# Nanoscale Strain Evolution and Grain Boundary-Mediated Defect Sink Behavior in Irradiated SiC: Insights from N-PED and DFT

N. Daghbouj[a*], A.T. AlMotasem[a*], J.Duchoň[b], B.S. Li[c], M. Bensalem[a], F. Munnik [d], Xin Ou[e], A. Macková[f], W.J. Weber[g], T. Polcar[a,h]

[a]*Department of Control Engineering, Faculty of Electrical Engineering, Czech Technical University in Prague, Technická 2, 160 00 Prague 6, Czechia*

[b]*Institute of Physics of the Czech Academy of Sciences, Na Slovance 1999/2, 182 21 Prague 8, Czechia*

[c]*State Key Laboratory for Environment-friendly Energy Materials, Southwest University and Technology, Mianyang, Sichuan 621010, China*

[d]*Helmholtz–Zentrum Dresden–Rossendorf, Institute of Ion Beam Physics and Materials Research, Bautzner Landstr. 400, 01328 Dresden, Germany*

[e]*State Key Laboratory of Functional Materials for Informatics, Shanghai Institute of Microsystem and Information Technology, Chinese Academy of Sciences, 865 Changning Road, Shanghai 200050, China*

[f]*Nuclear Physics Institute of the Czech Academy of Sciences, 250 68 Husinec-Řež, Czechia*

[g]*Department of Materials Science & Engineering, University of Tennessee, Knoxville, TN 37996, USA*

[h]*School of Engineering, University of Southampton, Southampton SO17 1BJ, United Kingdom*

**Abstract**

Understanding irradiation-induced strain in silicon carbide (SiC) is essential for designing radiation-tolerant ceramic materials. However, conventional methods often fail to resolve nanoscale strain gradients, especially in polycrystalline forms. In this study, we employ nano-beam precession electron diffraction (N-PED) to perform high-resolution, multi-directional strain mapping in both single-crystal 4H-SiC and polycrystalline α-SiC subjected to helium and hydrogen ion irradiation. The high-resolution X-ray diffraction (HR-XRD) simulations of He + H irradiated single-crystal 4H-SiC closely match the strain profiles obtained from N-PED, demonstrating the reliability and accuracy of the N-PED method. In He-irradiated polycrystalline α-SiC at high temperatures, a bubble-depleted zone (BDZ) near the grain boundary (GB) reveals that GBs act as active sinks for irradiation-induced defects. N-PED further shows strain amplification localized at the GBs, reaching up to ~2.5%, along with strain relief within the BDZ. To explain this behavior, density functional theory (DFT) calculations of binding and migration energies indicate a strong tendency for Si, C, and He atoms to segregate toward the GB core. This segregation reduces the availability of vacancies to accommodate He atoms and leads to local strain relaxation near the GB. Furthermore, first-principles tensile simulations reveal that Si and C interstitials mitigate He-induced GB embrittlement. Charge density and DOS analyses link this effect to the bonding characteristics between point defects and neighboring atoms at GB. These insights underscore the importance of grain boundary engineering in enhancing radiation tolerance of SiC for nuclear and space applications.

* Corresponding author's e-mail: daghbnab@fel.cvut.cz,  alasqahm@fel.cvut.cz

# 1. Introduction

Understanding the formation and evolution of strain in ion-irradiated materials is crucial for the development of radiation-resistant materials. Various experimental techniques, including Rutherford backscattering spectrometry in channeling mode, Raman spectroscopy, transmission electron microscopy (TEM), and Doppler broadening of positron-electron annihilation peaks, have been employed to detect and quantify irradiation-induced defects [1–4]. Irradiation generates correlated atomic displacements, inducing elastic strain that provides essential insights into defect characteristics and concentrations [5–7]. Despite its significance, elastic strain fields resulting from irradiation-induced defects remain underexplored, even though elastic interactions strongly influence defect evolution and growth [8,9]. These defects drive critical changes in material properties, including hardening, blistering, and embrittlement, which are detrimental to material performance in extreme environments [10–13]. Surprisingly, despite their relevance, the mechanisms underlying strain formation and evolution in irradiated materials are still not fully understood.

X-ray diffraction (XRD) is a fundamental technique for measuring irradiation-induced strain due to its sensitivity to atomic displacement fields. In single-crystal materials, strain distribution can be determined using high-resolution XRD combined with simulations, as demonstrated in prior studies [6,14,15]. At high ion fluences, strain saturation phenomena are often observed [16], typically attributed to the equilibrium between interstitial and vacancy-induced strain [17]. However, the complexity of XRD analysis at high fluences may contribute to the assumption of saturation. Furthermore, in polycrystalline materials, which are commonly used in reactor environments, XRD can only provide averaged strain values, lacking localized or depth-resolved information. Although XRD remains a powerful analytical tool, its inability to resolve nanoscale strain fields, particularly in polycrystalline materials, limits its effectiveness in predicting radiation-induced degradation.

Several advanced experimental approaches have been developed to overcome these limitations. Synchrotron X-ray micro-Laue diffraction enables non-invasive, depth-resolved strain measurements at micrometer scales [18,19]. Multi-reflection Bragg coherent X-ray diffraction imaging (MBCDI) has achieved ~30 nm 3D resolution of strain fields in ion-irradiated materials [20–22], revealing nanoscale strain fluctuations in self-ion-irradiated tungsten [20]. However, MBCDI requires complex sample preparation and synchrotron access, limiting its widespread use. High-angular-resolution electron backscatter diffraction (HR-EBSD) provides rapid, non-destructive surface strain mapping [18]. In self-ion-irradiated tungsten, HR-EBSD successfully captured long-range strain fluctuations at 1 displacements per atom (dpa) due to defect self-organization but lacked sufficient resolution to detect short-range distortions at lower doses (0.01 dpa) [19]. Other promising methods, such as high-resolution transmission Kikuchi diffraction (HR-TKD), with a spatial resolution of 5–10 nm [21–23], and four-dimensional scanning TEM (4D-STEM), which achieves 1 nm resolution[24], enable strain field mapping around individual dislocations.

To overcome the limitations of conventional strain measurement methods such as XRD, particularly their inability to resolve nanoscale variations and grain boundary effects, this study employs Nano-Beam Precession Electron Diffraction (N-PED). This technique enables high-resolution, spatially resolved strain mapping in both single-crystal and polycrystalline SiC subjected to ion irradiation. Direct comparison with high-resolution XRD validates the accuracy of N-PED, while the ASTAR strain mapping reveals localized elastic strain distributions, especially at GBs. By correlating these localized strain fields with defect accumulation, we provide critical insight into the mechanisms of irradiation-induced swelling and embrittlement. These findings enhance our ability to assess mechanical degradation in SiC, a key candidate material for next-generation nuclear reactors [25]. Moreover, it is well established that GBs play a crucial role as defect sinks and in the formation of defect-denuded zones adjacent to GBs [26–31]. The sink efficiency of interfaces or GBs under irradiation is not static but evolves dynamically with the accumulation of defects within the boundary region. This behavior is influenced by the specific character and atomic structure of the GB, as well as by nonequilibrium irradiation conditions [32–34]. Localized stress fields near GBs have been shown, through Monte Carlo simulations, to significantly impact elastic interactions with point defects. These stresses can enhance the interfacial sink strength by lowering the migration barriers for defect diffusion [35]. In addition, inert gases such as helium (He), xenon (Xe), and krypton (Kr), which exhibit very low solubility in solids, tend to migrate toward GBs where they can accumulate and form gas bubbles. These bubbles may compromise the structural integrity and fuel performance in nuclear environments [36]. The structure and properties of the interface or GB are therefore critical in governing not only bubble nucleation but also the transport and potential release of fission gases. Studies have shown that the diffusivity of gas atoms is typically higher along GBs than in the bulk, which facilitates preferential gas accumulation at the boundaries. Despite extensive work on defect transport and bubble formation at GBs, relatively few studies have investigated the strain evolution associated with defect sinks or in defect-denuded zones near GBs [30]. To our knowledge, there has been no detailed investigation into how GBs accommodate point defect sinks in relation to their mechanical integrity. Grain boundaries in polycrystalline SiC act as preferential sinks for irradiation-induced point defects, which can significantly influence the strain distribution and thus the mechanical strength of the material. Understanding how these defect-GB interactions affect nanoscale strain evolution is crucial for developing radiation-tolerant ceramics.

In this study, we address this gap by examining the local strain distribution near GBs in α-SiC after high-temperature helium irradiation. Additionally, we assess the fracture strength of GBs in the presence of point defects, providing new insights into the interplay between irradiation-induced damage and interfacial mechanical stability.

## 2. Experimental Section and Computational Methodology

### 2.1 Experimental Setup

### 2.1.1 Ion irradiation

Single-crystalline 4H-SiC samples were irradiated with He ions at a fluence of $1\times10^{16}$ He/cm² using 130 keV energy. Following irradiation, the samples were annealed at 800 °C, then co-irradiated with H ions at the same fluence ($1\times10^{16}$ H/cm²) using 110 keV. A separate polycrystalline α-SiC sample (120 μm thick) was irradiated with 300 keV He ions at a fluence of $1\times10^{17}$ He/cm² at 800 °C under a vacuum of a few millibars. The irradiation experiment took into consideration the in-depth ranges achieved by both ion species along the depth of the SiC sample as calculated by the SRIM code [37]. Corresponding irradiation damages induced by single He and H implantation in pure SiC were also evaluated by the SRIM code with the full-cascade simulations [38], a displacement energy of 20 eV, and 35 eV for C and Si [39], respectively, with an atomic density of 3.21 g/cm³.

### 2.1.2 Elastic Recoil Detection Analysis (ERDA)

To determine the depth profiles of He and H in the samples, Elastic Recoil Detection Analysis (ERDA) was performed using a 43 MeV $Cl^{7+}$ ion beam. The beam was incident at an angle of 75° with respect to the sample normal, and the scattering angle was set to 30°. The analyzed area covered approximately 2 × 2 mm². Recoil atoms and scattered ions were detected using a Bragg Ionization Chamber (BIC), which allows for simultaneous measurement of particle energy and atomic number (Z) identification. A separate solid-state detector was employed specifically for detecting H and He recoils. This detector was positioned at a scattering angle of 40° and shielded by a 25 µm Kapton foil, which blocks scattered ions and heavier recoils. However, the presence of the Kapton foil introduces energy loss straggling, reducing the depth resolution of the system. The ion beam dose (reported in arbitrary units) was monitored using a gold-coated rotating vane operating at 1 Hz, in conjunction with a solid-state detector that measured Cl ions backscattered from the gold surface. Data from the BIC, the H/He detector, and the rotating vane were recorded in list mode, enabling time-resolved analysis of elemental changes during measurement, especially the loss of volatile species such as H due to ion beam exposure. To quantify elemental loss, particularly of H and He, the list-mode data were used to plot total recoil counts as a function of ion dose. These plots were then fitted and extrapolated to zero dose to estimate the original elemental concentrations prior to any ion-induced depletion. The collected data were analyzed using NDF software version 9.6i [40]. ERDA spectra for carbon (C), oxygen (O), silicon (Si), and the combined H+He signal, as well as the Cl backscattering spectrum from Si, were extracted from the list-mode files. These spectra were simultaneously fitted using the NDF software. During analysis, a physical model of the sample was constructed to simulate expected spectra. These simulated spectra were iteratively compared and fitted to the experimental data. The final results include the best-fit spectra, an optimized sample model, and quantitative depth profiles of elemental concentrations derived directly from the experimental measurements.

### 2.1.3 Rutherford Backscattering Spectrometry in Channeling mode (RBS-C)

Using ion beam analysis techniques, specifically Rutherford Backscattering Spectrometry in Channeling mode (RBS-C), at the Tandetron MC 4130 tandem accelerator located at the Nuclear Physics Institute in Řež, near Prague, Czech Republic. For the RBS experiments, a beam of $He^+$ ions with an energy of 2.0 MeV was employed. The ions backscattered from the implanted samples were detected using an ORTEC ULTRA-series silicon detector, featuring an active area of 50 mm² and a depletion layer thickness of 300 µm. The detector was positioned at a backscattering angle of 170°, outside the scattering plane (Cornell geometry), to optimize energy resolution and minimize background noise. Energy calibration of the detection system was performed with a channel-to-energy conversion factor of 2.33813 keV/channel and a calibration offset of 91.015 keV. The overall energy resolution of the detector system was measured to be approximately 25 keV, ensuring accurate identification of elemental signals and defect-related features.

### 2.1.4 High Resolution X-ray diffraction (HRXRD)

Post-irradiation, X-ray diffraction (XRD) analysis was carried out using a RIGAKU 9 diffractometer equipped with a Cu K$\alpha_1$ X-ray source ($\lambda$ = 1.5406 Å), a parabolic multilayer mirror, and a two-reflection asymmetrically cut Ge (220) monochromator. The system included a linear position-sensitive detector covering a 2° 2θ range with 0.01° resolution. Symmetrical high-resolution θ-2θ scans were performed by selecting a single detector channel. The 0004 reflection of 4H-SiC (2θ$_B$ = 41.394°) was scanned in 0.005° steps over a wide enough range to capture all diffraction signals from the irradiated region. These scans enabled the determination of the lattice strain gradient normal to the sample surface [9,16,41]. Simulations of the XRD curves were performed using a combination of ion implantation and XRD modeling codes [42].

### 2.1.5 Strain Mapping by Nano-Precession Electron Diffraction (N-PED)

Strain mapping was performed using an FEI Tecnai TF20 transmission electron microscope (TEM) operated at 200 kV, equipped with a NanoMEGAS DigiSTAR precession device and controlled via Topspin software (NanoMEGAS) [2,43,44]. The microscope used a 20 µm condenser aperture to generate a nearly parallel beam, producing sharp spot diffraction patterns instead of disks. The method utilized nano-precession electron diffraction (N-PED), where a precessed electron beam was scanned across the region of interest with a step size of 5 nm and a beam size of 4 nm. Diffraction patterns (DPs) were collected at each location using a 1° precession angle and an exposure time of approximately 20 ms. A high-resolution scan was conducted to capture the strain field across the irradiated area, including grain GBs. Strain and orientation maps were generated from the DPs using Topspin software, which applies a cross-correlation algorithm to estimate diffraction spot positions with sub-pixel accuracy. The 2D elastic strain fields were calculated by comparing each DP to a reference pattern taken from a nominally strain-free, unirradiated region of the sample ($\varepsilon_{zz}$ = 0) [45]. This reference point was carefully selected to avoid irradiation damage and accurately represent the undistorted lattice. The NanoMEGAS ASTAR system employs precession electron diffraction and automated indexing to produce orientation and phase maps with nanometer-scale spatial resolution [46]. In collaboration with AppFive, NanoMEGAS developed the "AutoSTRAIN acquisition & analysis module," which enables high-precision elastic strain mapping. The Topspin software automates beam precession, scanning, and DP acquisition and analysis[47,48]. Beam precession significantly reduces artifacts due to dynamical scattering and thickness variations, while also increasing the number of observable diffraction spots in each pattern, enhancing accuracy and spatial resolution of the strain maps [49–51]. Virtual bright-field scanning transmission electron microscope (STEM) images were generated by placing a digital aperture over the central diffraction spot, revealing contrast in the scanned region.

## 2.2 DFT calculations

Density Functional Theory (DFT) calculations were conducted using the projector augmented wave (PAW) method [52] and the Perdew-Burke-Ernzerhof functional for solids (PBE) [53], as implemented in the Vienna Ab Initio Simulation Package (VASP) [54]. The PAW potentials for silicon (Si) and carbon (C) included 4 valence electrons each. The 6H-SiC crystal adopts a hexagonal close-packed (hcp) structure with space group $P6_3mc$ and lattice parameters $a$ = 3.08 Å, $b$ = 3.08 Å, and $c$ = 15.10 Å. For the single-crystal calculations, a supercell containing 192 atoms was constructed, comprising 4 × 4 × 1 unit cells aligned along the $x$-, $y$-, and $z$-directions, respectively. The simulation box is oriented in the $x$ $[11\bar{2}0]$, $y$ $[1\bar{1}00]$, and $z$ $[0001]$ directions. The symmetric tilt grain boundary bicrystal (STGB) of 6H-SiC was constructed using Atomsk [55] following the procedures described in Ref.[56]. Here, we selected an STGB with a misorientation angle of θ = 73.74° as a representative of the high-angle GB class. This type of boundary is characterized by a large excess free volume at its core, which can accommodate multiple interstitial atoms and thereby qualitatively reproduces the features observed in TEM experiments. The STGB was created by partitioning the supercell into two halves, followed by a rigid-body rotation of the upper half by an angle θ/2 (clockwise) and the lower half by an angle -θ/2 (counterclockwise) about the $x$ $[11\bar{2}0]$ axis, where the total misorientation angle was θ = 73.74°. This geometric construction ensures that the two crystals share a common tilt axis while maintaining mirror symmetry across the grain boundary plane, a characteristic feature of STGBs that facilitates the study of their structural and energetic properties. The dimensions of (STGB) 6H-SiC are 6.19 Å, 20.93 Å, and 19.72

Å 228 atoms, see Fig. 9b. The constructed grain boundary (GB) model was first relaxed using molecular statics (MS) within the γ-surface framework [57]. This approach involves displacing the two crystals relative to each other to explore various initial configurations. Each configuration is then relaxed to identify potential metastable states. The interaction between Si and C atoms was described using the interatomic potential from [58]. The resulting structure then undergoes further structural relaxation using the conjugate gradient method within the (DFT). The climbing image nudged elastic band (CI-NEB) method, as implemented in the VTST tools [59], was employed to calculate the migration energy of atoms. A minimum of nine images along the migration path was used to ensure convergence of the migration barrier. All supercells were fully relaxed, atomic coordinates and volume were optimized until the forces on all ions were reduced below 0.01 eV/Å. A plane-wave energy cutoff of 450 eV was used for all calculations. Brillouin zone integrations were performed using a 5 × 5 × 5 $k$-point mesh for single-crystal structures and a 4 × 2 × 2 mesh for the STGB models using the Monkhorst-Pack method [60]. A uniaxial tensile strain of 0.02 was applied along the [1000] direction, perpendicular to the grain boundary (GB), with atomic motion constrained in the lateral directions. At each strain increment, the atomic configuration was derived by uniformly scaling the fully relaxed structure from the preceding step, maintaining a continuous and consistent strain progression.

## 3. Results and discussion

### 3.1. Ions and damage distributions

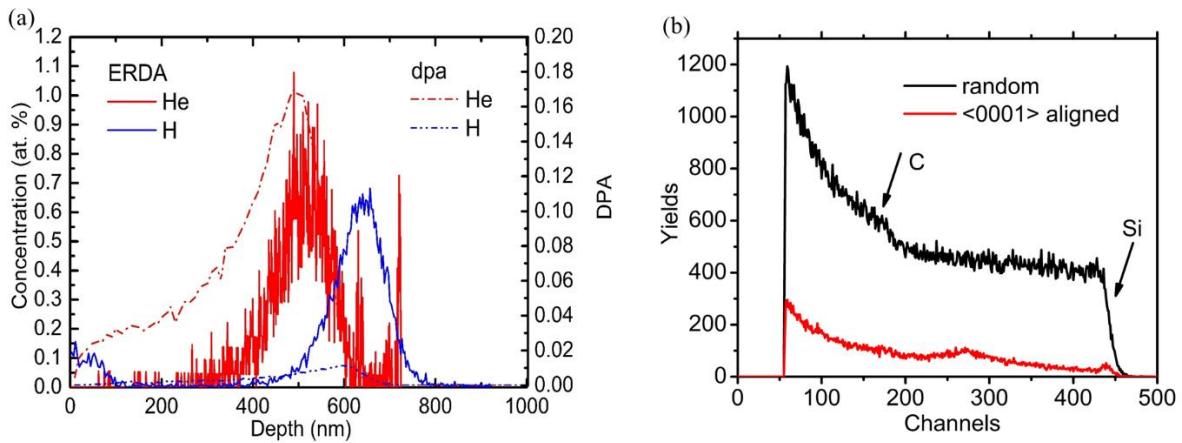

**Fig. 1.** (a) ERDA spectra for the 4H-SiC sample after sequential helium (He) and hydrogen (H) ion irradiation with 1 x $10^{16}$ ions/cm$^2$, along with the corresponding displacement damage profile

simulated using SRIM, (b) RBS-C spectra in random and <0001>-aligned configurations, highlighting Si and C edges.

ERDA was performed to determine the depth distribution of He and H in the irradiated 4H-SiC sample, as shown in Fig. 1. These experimentally measured concentration profiles are compared with SRIM simulations in Fig. S1. The measured profiles for both He and H closely align with the SRIM-predicted distributions. Post-irradiation annealing at 800 °C, performed after He implantation, does not significantly alter the He distribution. This suggests that the He atoms are retained in their implanted positions and do not diffuse substantially at this annealing temperature. Additionally, the subsequent H implantation appears unaffected by pre-existing He atoms or He-induced damage, indicating minimal interaction between the two ion species during irradiation. The damage profile, expressed in DPA and also simulated using SRIM, shows a slightly shallower distribution compared to the ion concentration profiles. This is expected since damage tends to peak slightly before the ion's projected range. As observed in the figure, the projected range (Rp) of H and He ions follows a Gaussian-like distribution, with the peak positions located at approximately 680 nm and 540 nm, respectively. A small H concentration is detected near the surface (depth ⩽ 100 nm), which is attributed to surface contamination, likely due to exposure to air and lack of vacuum storage. These surface values are therefore excluded from the discussion of intrinsic H implantation behavior.

This sample was characterized using the Rutherford Backscattering Spectrometry in Channeling mode (RBS-C) technique. Representative RBS-C spectra are presented in Fig. 1b, where the signal edges corresponding to Si and C are marked. When the incident ion beam is aligned along a major crystallographic axis, such as the <0001> direction in single-crystal SiC, the backscattered ion yield is significantly reduced compared to a random orientation. This occurs because the ions are channeled along open atomic rows or planes, reducing the likelihood of direct collisions with nuclei and thereby lowering the backscattering yield (as shown in Fig. 1b). This reduction is sensitive to lattice quality, making RBS-C a powerful method to detect radiation-induced lattice disorder and point defects.

To quantitatively assess the level of crystalline damage, we use the normalized yield, $\chi$, defined as the ratio of the backscattering yield in the irradiated (channeled) sample to that in a random or completely amorphous reference sample, calculated over a specific region of interest (ROI) in the spectrum. In this study, the ROI spans channels ~230–320, corresponding to the depth distribution of the implanted ions and the associated damage region in the SiC lattice. The key metric derived from RBS-C is the minimum channeling yield ($\chi\_min$), which reflects the degree of disorder in the crystal lattice. For the irradiated sample, a $\chi\_min$ value of 19.1% was obtained in the damage zone (channels 230–320), indicating lattice disruption relative to the virgin SiC substrate. In contrast, a defect-free single crystal would exhibit a much lower $\chi\_min$ (~1–3%). Therefore, the elevated value confirms the presence of He-induced defects such as interstitials and interstitial-type defects.

## 3. 2 Depth-resolved nanoscale strain in He and H co-implanted single crystal 4H-SiC

### 3. 2. 1 HRXRD

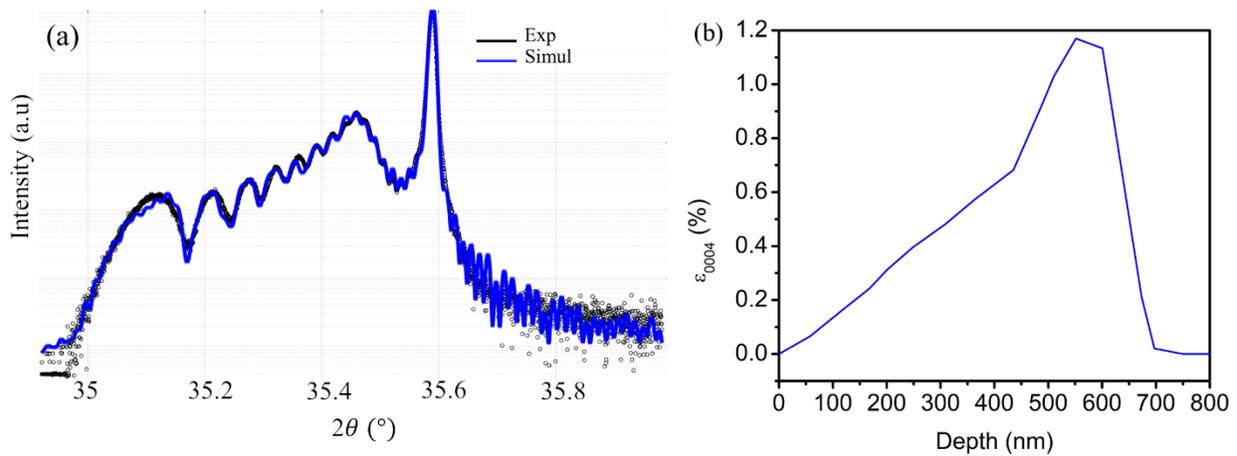

**Fig. 2.** (a) High-resolution X-ray diffraction (HR-XRD) spectrum (black: experimental; blue: simulation) measured near the symmetric (0004) reflection of 4H–SiC after co-implantation with He and H ions at a fluence of $1 \times 10^{16}$ ions/cm². (b) Corresponding out-of-plane strain ($\varepsilon_{0004}$) profile as a function of depth, derived from fitting the experimental diffraction pattern using the dynamical theory of X-ray diffraction.

The HR-XRD measurements in Fig. 2(a) capture the structural modifications in 4H-SiC resulting from the co-implantation of He and H. The diffraction pattern reveals interference fringes on the low-angle side of the main (0004) Bragg peak of the SiC substrate, indicating the presence of a damaged layer with an expanded lattice parameter normal to the surface. These fringes arise due to constructive interference from regions with varying strain, which is induced by implanted ions and defect accumulation [8]. The simulated diffraction pattern, based on the dynamical theory of X-ray diffraction [42], shows excellent agreement with the experimental data, validating the assumed strain model. For the simulation, the damaged region is modeled as a stack of thin sublayers, each with a defined thickness and strain, where strain is assumed to scale proportionally with the local He and H concentrations. This modeling approach enables reconstruction of the strain depth profile, as shown in Fig. 2b. Fig. 2b presents the extracted out-of-plane strain ($\varepsilon_{0004}$) profile as a function of depth. The strain exhibits a Gaussian-like distribution, peaking at approximately 588 nm with a maximum strain of ~1.15%. This depth corresponds closely with the projected range of the He implanted ions, consistent with the ERDA and SRIM profiles shown in Fig. 1. The strain increases gradually from the surface, reaching its maximum in the high-damage region, and then sharply decreases beyond ~700 nm. This indicates that lattice distortion is localized within the ion-affected region, with negligible defect diffusion into the undamaged bulk of the substrate.

### 3. 2. 2 Nano-beam precession electron diffraction (N-PED) ASTAR strain mapping

Fig. 3 provides a comprehensive strain analysis of the He-H co-implanted 4H-SiC layer using TEM and nano-beam electron diffraction techniques. Panel (a) shows a cross-sectional BF-TEM image of the irradiated sample aligned along the [110] zone axis, chosen to visualize strain along the c-axis (0004 reflection), which is perpendicular to the sample surface. The sum of He + H damage profile (magenta line) calculated via SRIM simulations is overlaid, indicating that the peak damage lies at ~570 nm. Panel (b) shows the selected area diffraction pattern (SADP) along the [110] zone axis, confirming the crystallographic orientation and allowing indexing of the relevant reflections used for strain analysis. Panel (c) indicates the direction along which the strain component $\varepsilon_{0004}$ (out-of-plane strain) is measured, based on shifts in the 0004 diffraction spots relative to a strain-free reference region. The reference region was taken from the unimplanted substrate, far from the irradiated layer, to ensure an accurate baseline. Panel (d) displays the spatial distribution of out-of-plane strain in the irradiated region as a 2D color map acquired via N-PED. The strain reaches a peak value around the projected range of the He implanted ions at ~ 588 nm and exhibits a nearly Gaussian distribution similar to that observed in the HR-XRD-derived profile. The high-strain region aligns well with the area of highest dpa. Panel (e) presents a quantitative comparison of the depth-dependent strain profile derived from N-PED (red curve) and HR-XRD simulation (blue curve). Both profiles exhibit similar trends, with peak strain values located around 588 nm. The maximum strain from TEM (~1.0%) is slightly lower than that from HR-XRD (~1.15%), likely due to relaxation effects introduced during TEM lamella preparation, such as thinning-induced strain relief or slight bending of the foil. We observe a measurable strain near the surface region (~0-60 nm), despite the low expected level of ion implantation damage there. We attribute this residual strain mainly to surface-related phenomena such as the formation of native oxide layers, which can introduce stress and slight lattice distortion, or to relaxation effects as the surface adjusts post-implantation. These surface-induced factors contribute to the observed strain even in areas with minimal direct displacement damage. While N-PED offers exceptional spatial resolution for strain mapping, it is susceptible to experimental challenges that can influence the accuracy of measurements. Beam-induced artifacts can modify local strain states, and the process of preparing thin TEM lamellae may cause partial relaxation of strain due to stress release, potentially lowering measured strain values. Furthermore, slight variations in electron beam precession angles and intensities may distort diffraction patterns, affecting precision. To improve statistical reliability, additional strain maps were acquired from larger areas (see Fig. S2 in the Supplementary Information), which confirmed the reproducibility of both the strain magnitude and distribution across different sample regions.

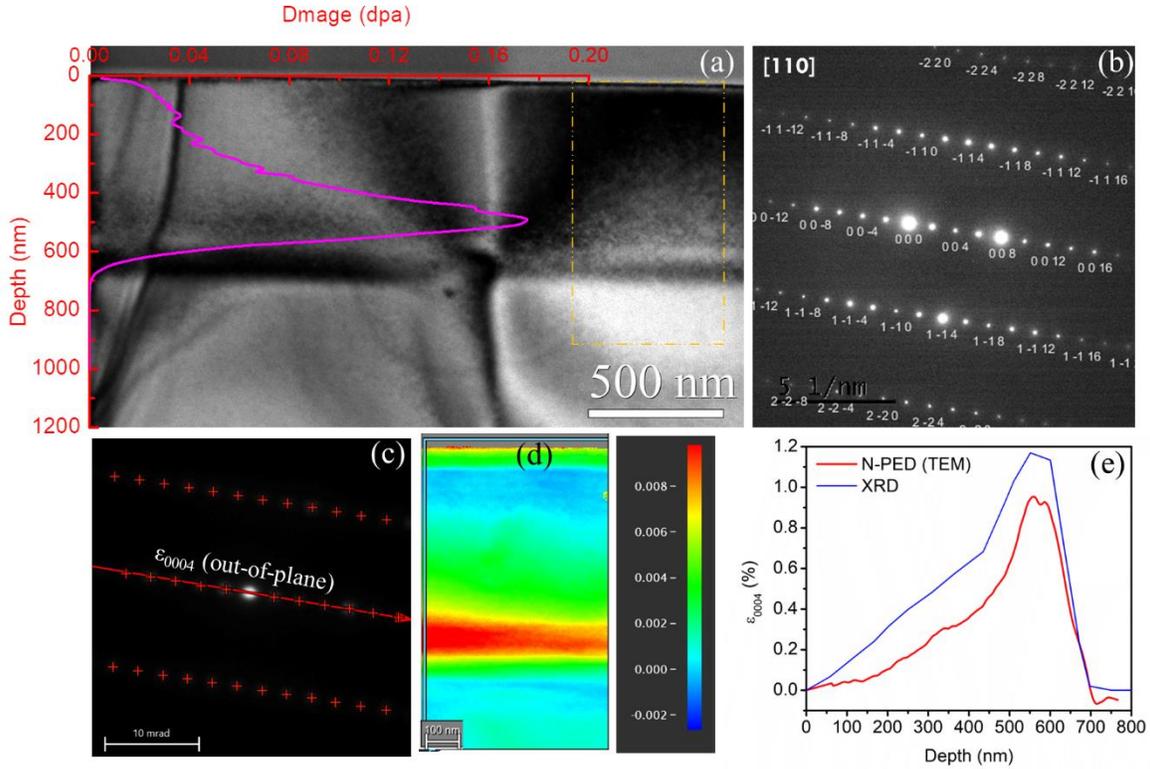

**Fig. 3.** Comprehensive strain analysis of He+H co-implanted 4H-SiC (fluence: $1 \times 10^{16}$ ions/cm² each). (a) Bright-field TEM image of the cross-section, with the SRIM-calculated displacement-per-atom (dpa) profile (magenta) superimposed. (b) Selected-area diffraction pattern confirming crystal orientation along [110]. (c) Schematic of the diffraction vector used for out-of-plane strain ($\varepsilon_{0004}$) using the 0004-diffraction spot. (d) Two-dimensional strain map of the out-of-plane component $\varepsilon_{0004}$ acquired using nano-beam precession electron diffraction (N-PED) in the region highlighted by the dashed square in (a), showing a Gaussian strain distribution localized at the He peak. (e) Quantitative comparison between HR-XRD (blue) and N-PED (red), confirming consistent peak strain values and depth distributions.

To achieve a comprehensive understanding of the strain distribution in ion-irradiated 4H-SiC, we analyzed strain components along multiple crystallographic directions using high-resolution N-PED. Fig. 4 illustrates the two-dimensional strain maps of the shear and in-plane components, providing insights into anisotropic deformation behavior.

Panel (a) shows a VBF-TEM image of the He-H co-implanted region, highlighting the irradiated zone. The top row [(b) and (c)] focuses on the shear strain $\varepsilon_{1\bar{1}0\bar{1}2}$ with panel (b) depicts the measurement geometry relative to the $1\bar{1}0\,\bar{1}2$ reflection. The corresponding strain map in panel (c) reveals heterogeneous shear distortions, with localized high-strain areas reaching up to ~0.6%. These distortions are primarily concentrated near the damage peak and suggest significant lattice rotation or distortion due to He-H interactions.

The bottom row [(d)–(f)] presents the in-plane strain $\varepsilon_{1\bar{1}00}$, with the diffraction vector aligned along the $[1100]$ direction. The strain map in panel (f) indicates a more uniformly distributed compressive strain field across the implanted region. These compressive features are likely the result of implantation-induced defects and volume contraction during relaxation.

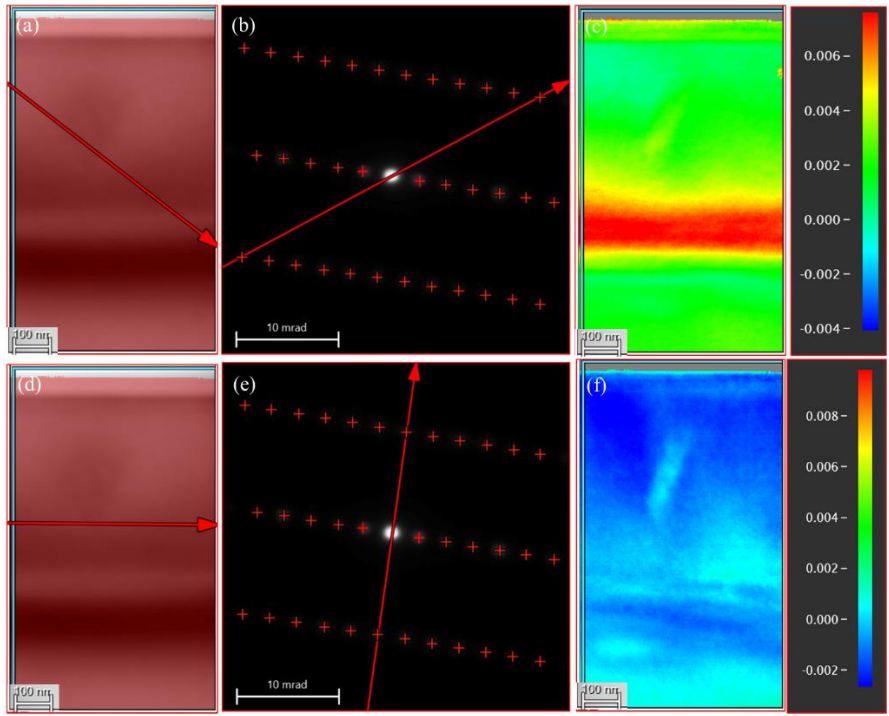

**Fig. 4.** (a) Virtual bright-field transmission electron microscopy (VBF-TEM) image of the He-H co-implanted 4H-SiC sample, oriented along the [110] zone axis. Strain Analysis (Top Row): (b) Schematic showing the measurement direction for the strain component $\varepsilon_{1\bar{1}0\,\bar{1}2}$, derived from the $1\bar{1}0\bar{1}2$ diffraction spot in the nano-beam precession electron diffraction (N-PED) pattern. (c) N-

PED strain map corresponding to $\varepsilon_{1\bar{1}0\bar{1}2}$ for the region indicated in Fig. 3 (a), revealing tensile distortions. In-Plane Strain Analysis (Bottom Row): (d) Schematic indicating the measurement direction for the in-plane strain component $\varepsilon_{1\bar{1}00}$, using the $1\bar{1}00$ diffraction spot. (e) Corresponding N-PED strain map for $\varepsilon_{1\bar{1}00}$, showing lateral compressive strain throughout the irradiated region. (f) Strain distribution with a color scale indicating strain values ranging from tensile (red/yellow) to compressive (blue).

To gain a more comprehensive understanding of the strain state in the ion-irradiated 4H-SiC, we extended our analysis beyond the purely out-of-plane strain component. Fig. 5 presents depth-dependent strain profiles derived from N-PED measurements along various crystallographic directions, and the details of strain mapping are shown in Fig. S3. These directions are $\varepsilon_{1\bar{1}0x}$ directions (e.g., $\varepsilon_{1\bar{1}0\bar{2}}$, $\varepsilon_{1\bar{1}0\bar{4}}$, $\varepsilon_{1\bar{1}0\bar{6}}$, $\varepsilon_{1\bar{1}0\overline{10}}$, $\varepsilon_{1\bar{1}0\overline{12}}$): intermediate directions, combining in-plane and out-of-plane character. These oblique components are especially valuable for assessing anisotropic strain and understanding the coupling between axial and lateral deformation under irradiation.

The profiles confirm a peak strain around 588 nm, consistent with the He peak concentration, but show different magnitudes and depth distributions across directions. The out-of-plane strain $\varepsilon_{0004}$ exhibits the highest peak (~1%), while oblique and in-plane components show lower amplitudes. This disparity reflects the anisotropic elastic and damage response of 4H-SiC, governed by its hexagonal crystal structure and the preferential alignment of implanted defects along the c-axis. The elevated strain in the (0004) direction arises because implanted He and H defects, including interstitials, bubbles, and platelets, tend to expand the lattice normal to the surface, creating tensile out-of-plane strain. In contrast, the surrounding matrix constrains lateral expansion, leading to compressive in-plane stress. This axial-lateral coupling is especially pronounced in anisotropic materials like 4H-SiC, and more details will be presented in section 2.4.

This variation underscores that the strain is highly anisotropic, with the degree of lattice distortion depending strongly on crystallographic orientation. The inclusion of oblique directions provides critical insight into the 3D strain state, revealing complex relaxation behavior that would be missed if only pure in-plane and out-of-plane directions were considered. The differences in strain magnitude and spatial extent among the components highlight the anisotropic elastic response of 4H-SiC under irradiation. Notably, the steeper gradient in $\varepsilon_{0004}$ compared to the in-plane strains suggests differential relaxation mechanisms, as well as more localized defect accumulation and relaxation along the c-axis. Since strain directly influences defect migration, coalescence, and trapping, especially at interfaces like GBs, quantifying strain anisotropy is essential for predicting material degradation pathways. For nuclear-grade SiC, this approach offers key insights into mechanical stability, swelling resistance, and long-term structural reliability under irradiation.

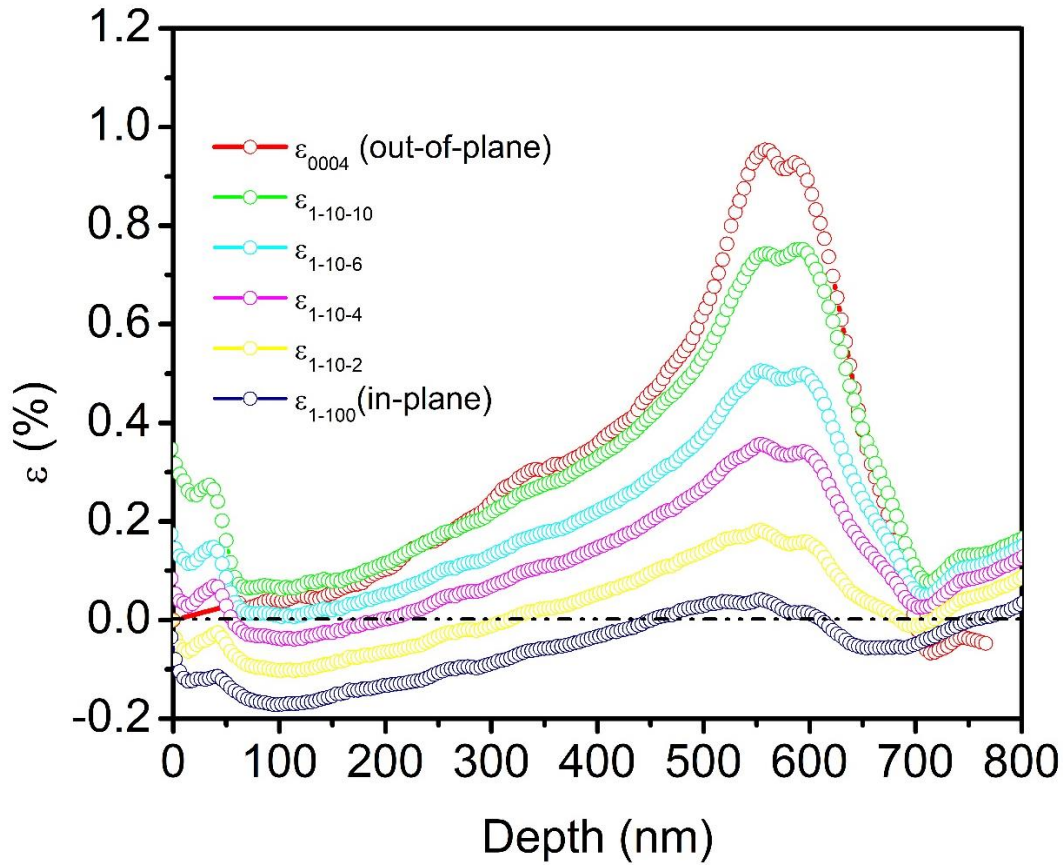

**Fig.5.** Depth-dependent strain profiles from N-PED in He–H co-implanted 4H-SiC. Multiple crystallographic directions are analyzed, including out-of-plane ($\varepsilon_{0004}$), in-plane ($\varepsilon_{1\bar{1}00}$) and multiple intermediate (oblique) directions such as $\varepsilon_{1\bar{1}0x}$, which combine in-plane and out-of-plane contributions. The maximum out-of-plane strain (~1%) occurs near the He projected range, while in-plane components exhibit compressive strain. The results demonstrate anisotropic strain evolution due to directional lattice expansion along the c-axis constrained by the SiC matrix.

## 3.3 Strain mapping in He irradiated polycrystalline SiC

Strain Mapping in He-Irradiated Polycrystalline 6H-SiC Using Nanoscale Diffraction Techniques
While strain mapping using HRXRD is well-established for single-crystal SiC, its application to polycrystalline SiC (poly-SiC), commonly used in nuclear environments, remains limited. This study addresses a critical knowledge gap by quantifying strain in He-irradiated poly-SiC, with particular focus on GBs strain evolution and radiation-induced damage effects.

Conventional XRD measurements revealed no significant diffraction peak shifts, suggesting an absence of measurable macroscopic strain, as shown in Fig. S4. However, XRD inherently provides depth-averaged, bulk information, which can mask localized strain gradients. In polycrystals, spatial heterogeneity in grain orientation and the superposition of opposing strain types (compressive vs. tensile) can lead to apparent strain cancellation. Furthermore, the deep penetration depth of X-rays exacerbates this averaging effect. Therefore, the lack of observed XRD peak shifts is likely due to the spatial averaging of localized, anisotropic strain distributions.

To overcome these limitations, we employed N-PED to probe strain at the nanoscale, specifically within the He implantation zone (see Fig. 6). This technique enables high-resolution, localized strain mapping in volumes significantly smaller than those probed by XRD. While XRD provides critical insights into strain in single-crystal materials, its spatial resolution typically exceeds several micrometers, making it ineffective for resolving nanoscale strain variations, particularly at GBs. This limitation restricts XRD's ability to capture localized strain amplification effects that could have significant implications for material stability. N-PED, with its sub-nm resolution, overcomes this constraint by enabling direct quantification of strain gradients at critical microstructural features, including GBs. The 300 keV He irradiation was conducted at 800 °C with a fluence of $1\times10^{17}$ He/cm², leading to the formation of helium-induced defects such as bubbles and platelets, which were predominantly located near the peak He concentration zone (Fig. 6a–b). Additional details of these He-related defects are provided in Fig. S5. Importantly, although GBs exhibited significant He accumulation, they retained their crystalline structure, as confirmed by high-resolution TEM images in Fig. S5e-f. These boundaries were surrounded by regions largely free of visible defects, forming so-called bubble-denuded zones (BDZs). This pattern supports the hypothesis that GBs act as efficient sinks for both He atoms and irradiation-induced point defects, leading to localized redistribution of damage [61,62]. To evaluate the resulting strain distribution, the sample was oriented along the [214] zone axis (Fig. 6d), which enables strain analysis in multiple crystallographic directions. We first examined the direction perpendicular to the GBs surface (i.e., the out-of-plane direction, along [0$\bar{1}\bar{1}$4] as shown in Fig. 6e-f. Nanoscale strain mapping revealed a localized tensile strain component in this direction, attributed to lattice expansion caused by He accumulation and defect clustering. The measured out-of-plane strain ($\varepsilon_\perp$) exhibited a Gaussian-like distribution, closely following the He implantation profile, and reached its maximum near the projected range of the implanted ions (Fig. 6g).

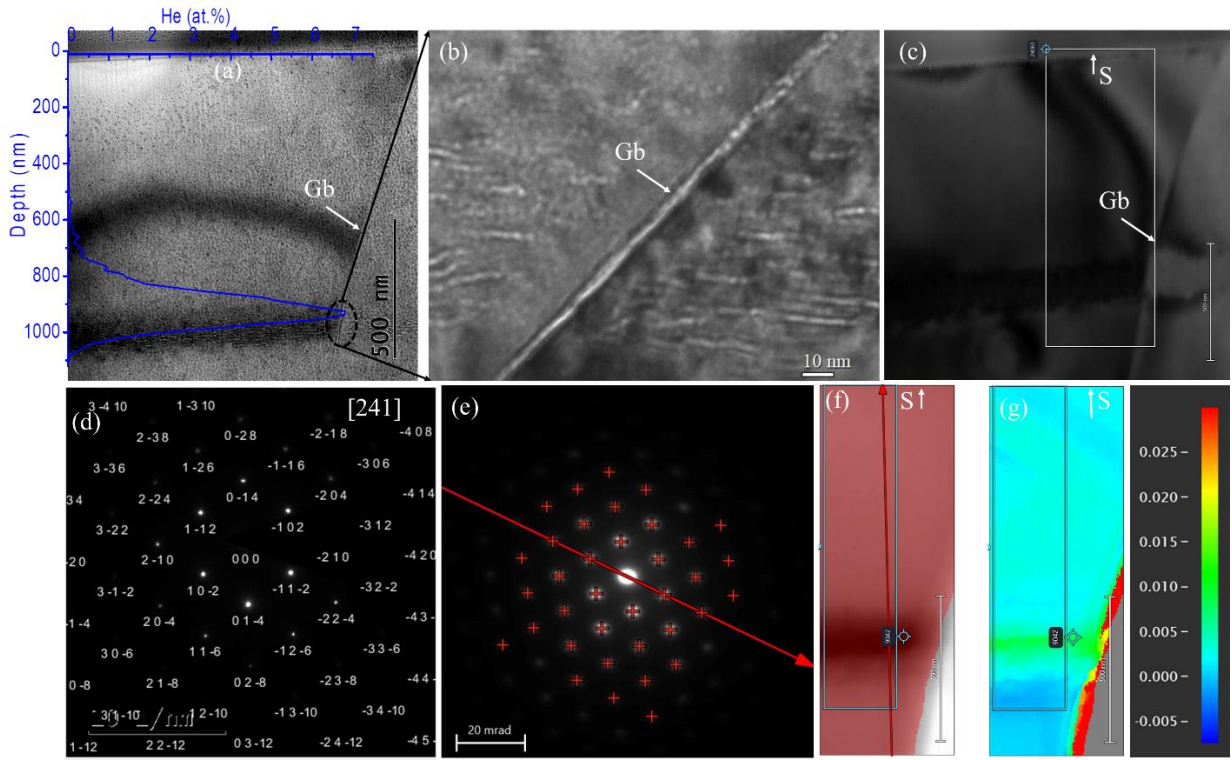

**Fig. 6.** Helium irradiation effects in polycrystalline SiC (800°C, 1×10$^{17}$ He/cm²). (a) Bright-field TEM image with the ERDA-simulated He profile (blue) superimposed. (b) High-magnification TEM showing He bubbles and platelets localized near the He concentration peak and grain boundary (Gb, dashed line). (c) Virtual STEM image marking the N-PED analysis region. (d) SAED pattern along [241]-zone axis SAED pattern. (e) direction of the measured strain in SADP

of $\varepsilon_{01\bar{1}\bar{4}}$ (out-of-plane) using the $[0\bar{1}\bar{1}4]$ diffraction spot. (f) VBF showing the direction of the strain with respect to the surface, (g) strain map of $\varepsilon_{01\bar{1}\bar{4}}$ revealing a Gaussian strain distribution.

Further, depth-resolved, multidirectional strain analysis using N-PED (Fig. 7) revealed compressive strains in in-plane ($\varepsilon_{2\bar{1}02}$) and near-in-plane directions, contrasting with the tensile out-of-plane strain. Additional mapping details are provided in Fig. S6. This anisotropic strain distribution reflects the directional nature of damage accumulation and mechanical constraint within the crystal. He-induced defects, such as bubbles and platelets, cause local lattice expansion. However, this expansion is constrained laterally by the surrounding crystalline matrix, which resists deformation in the in-plane directions. As a result, the material experiences biaxial compressive strain within the plane and tensile strain perpendicular to the surface.

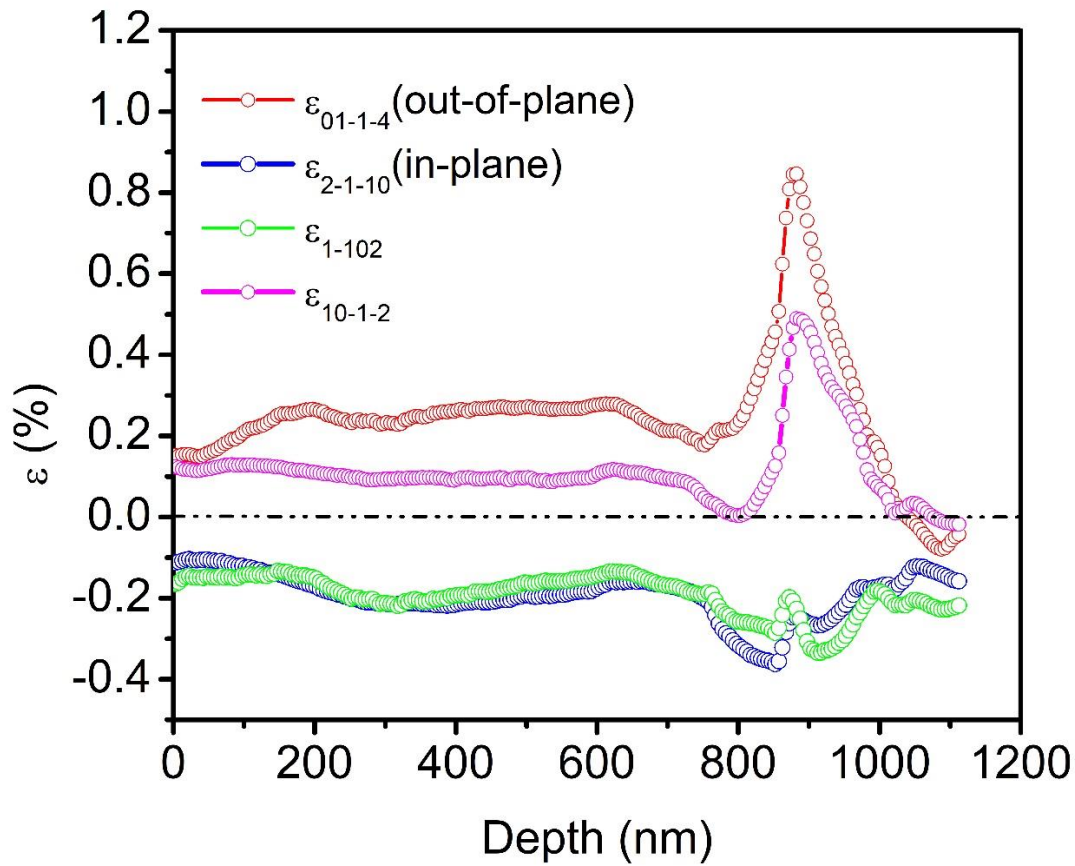

**Fig. 8.** Strain distribution near grain boundaries (GBs) in He-irradiated polycrystalline SiC. (a) Strain profile along highlighting tensile strain up to ~2.5% localized at GBs, with defect-denuded regions (dashed circles) $[0\bar{1}\bar{1}4]$. *(b)* Strain profile along $[1\bar{1}02]$ showing strain relief near GBs compared to the grain interior. Together, the results confirm that GBs act as efficient sinks, redistributing He and reducing local strain accumulation.

### 3.4 Grain Boundary-Mediated Strain Evolution and Defect Interaction in Irradiated SiC

Radiation tolerance in crystalline materials like SiC critically depends on how effectively GBs act as sinks for irradiation-induced defects. One key indicator of a GB's defect-absorbing efficiency is the formation of a denuded zone, a region adjacent to the GB that is largely free of defects such as He bubbles or platelets. This feature, visible in Fig. 6b, reflects the GB's capacity to attract and absorb point defects and He atoms, thereby mitigating damage accumulation in the surrounding matrix. The strain around twins, dislocations, and GBs has been performed in other studies [63–66], and the evolution of strain along and around GBs is shown in Fig. 8. Interestingly, tensile strain as high as ~2.5% is observed along the GB, even in regions far from the projected He concentration peak. This elevated strain is primarily attributed to lattice mismatch and structural discontinuities at the GBs, which disrupt lattice coherence and introduce localized mechanical stresses.

In contrast, strain relief is evident in areas adjacent to the GB near the He concentration peak, as highlighted by the dashed lines in Fig. 8a, b. This localized relaxation is particularly noticeable in the in-plane direction. While one might expect increased strain near GBs due to defect accumulation, the formation of a bubble-depleted (denuded) region suggests active defect recombination or diffusion toward the boundary. In essence, GBs serve as efficient sinks not only for interstitials and vacancies but also for He, reducing the local defect pressure and thus lowering strain in nearby regions [61,67–69]. However, the distribution of strain is highly anisotropic, with certain regions under tensile stress and others under compression. Such uneven stress fields, caused by defect-driven strain localization and directional relaxation, may act as precursors to microcrack initiation or intergranular fracture. Therefore, while GBs improve radiation tolerance by acting as defect sinks, they also introduce complex strain fields that must be carefully understood when designing radiation-resistant polycrystalline ceramics.

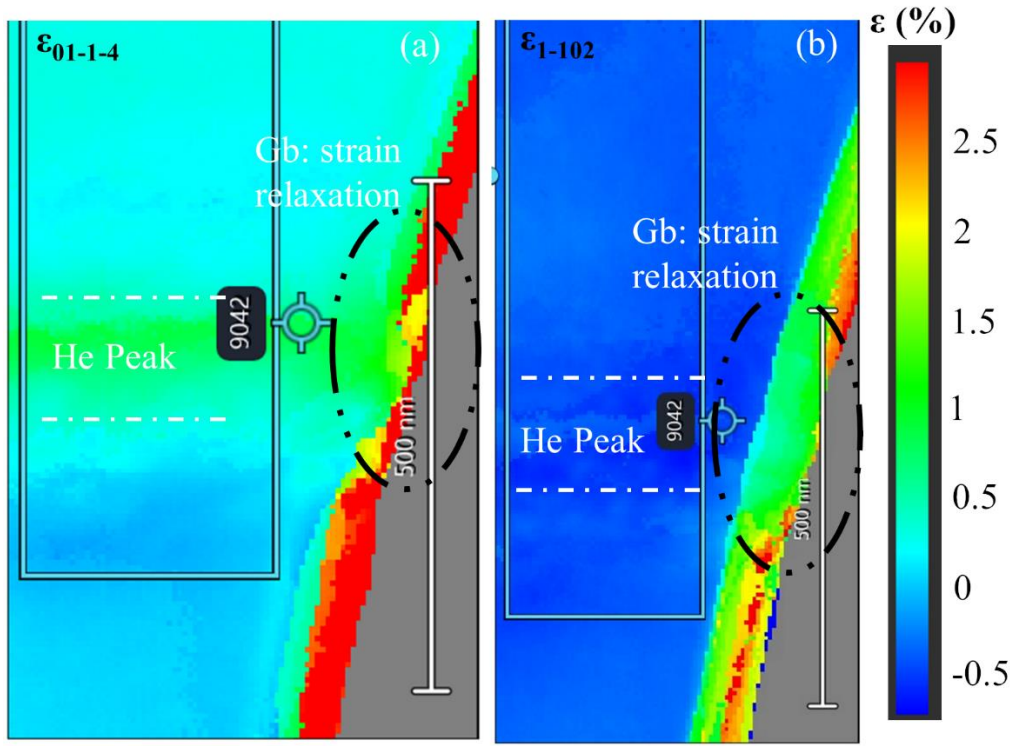

**Fig. 8.** Strain distribution near grain boundaries (GBs) in He-irradiated polycrystalline SiC. (a) Strain profile along highlighting tensile strain up to ~2.5% localized at GBs, with defect-denuded regions (dashed circles) $[0\bar{1}\bar{1}4]$. *(b)* Strain profile along $[1\bar{1}02]$ showing strain relief near GBs compared to the grain interior. Together, the results confirm that GBs act as efficient sinks, redistributing He and reducing local strain accumulation.

To validate this hypothesis, we performed density functional theory (DFT) calculations to determine the migration energy barriers of key point defects in SiC. A bicrystal model of 6H-SiC was constructed, incorporating STGB with a specific misorientation angle $\theta=73.74°$, as illustrated in Fig. 9b. Using the Nudged Elastic Band (NEB) method, we calculated the migration energy barriers for silicon, carbon, and He interstitials along various paths leading from atomic planes near the GBs (GB1 and GB2) toward the interface (Fig. 9c). The initial interstitial positions were chosen at the R sites, which are identified as the most energetically favorable locations (highlighted in red in Fig. 9a). The results show that silicon interstitials exhibit low migration barriers near the GB, less than 0.357 eV within the three atomic layers closest to the interface, and approximately 1.26 eV in the second atomic plane. This suggests high defect mobility near the boundary, similar to previously observed behavior in 3C-SiC [70]. Carbon interstitials follow a comparable trend, with migration barriers of ~2.52 eV near the GB, increasing significantly to ~7.32 eV further away from the GB plane, indicating a strong tendency for segregation toward the boundary. He atoms also display a preferential migration toward GBs, consistent with our experimental observations of elevated He concentrations in GBs regions (Fig. 6b).

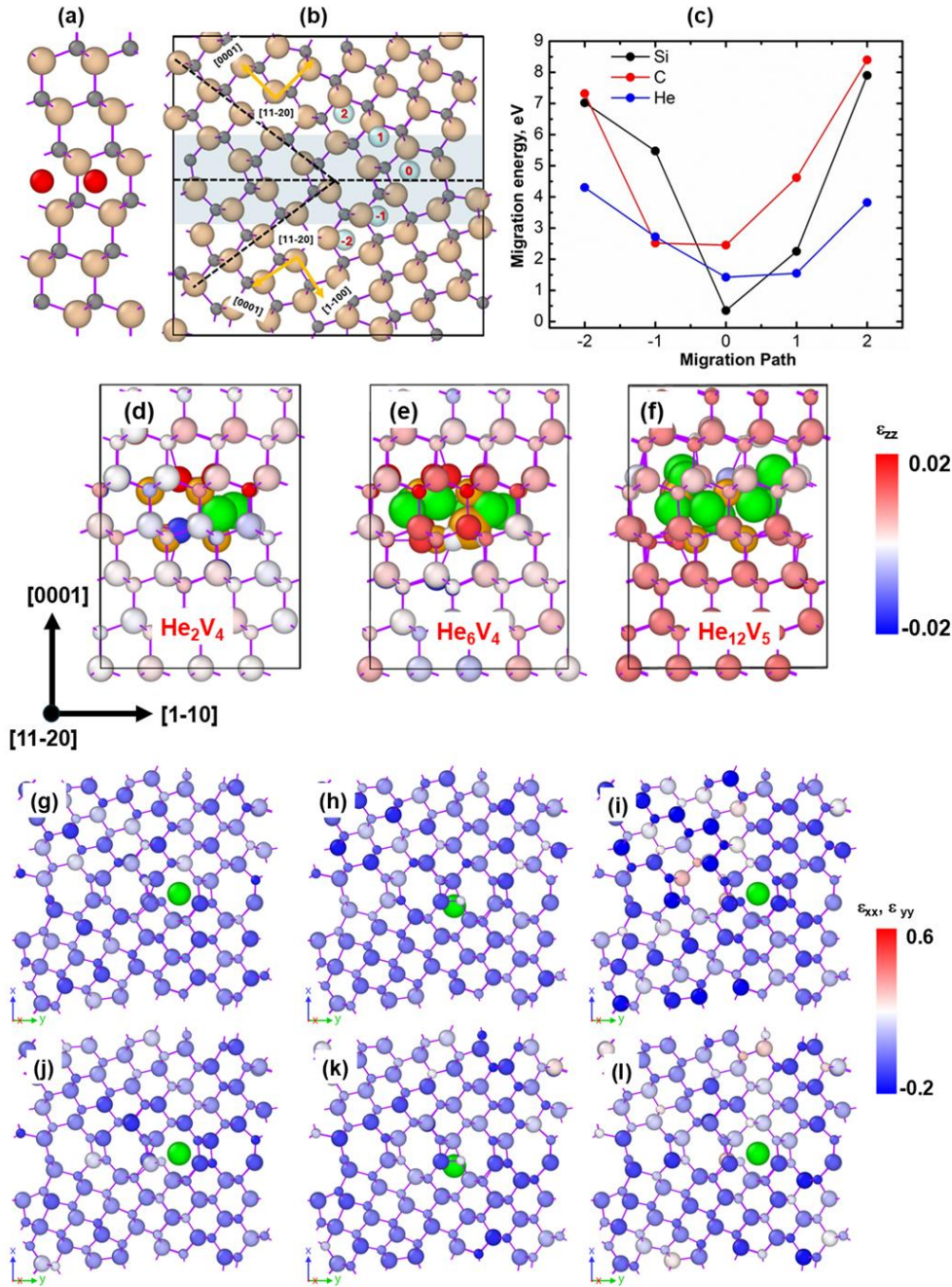

**Fig. 9**. (a) Atomic structure model of a conventional 2×2×1 supercell of 6H-SiC, red atoms shows the most energetically favorable site of interstitial (R site), (b) STGB model used for defect segregation studies. (c) Nudged Elastic Band (NEB) calculated migration energy barriers for silicon, carbon, and helium interstitials along the diffusion path highlighted by red numbers in Fig. 9 (b), illustrating reduced migration barriers in the vicinity of the grain boundary. Atomic snapshots showing the various defect configurations in 6H-SiC single crystal (d) Model of a smaller He-vacancy cluster ($He_2V_4$) composed of 4 carbon vacancies ($4V_C$) with a He/V ratio of 0.5, (e) smaller He-vacancy cluster ($He_6V_4$) composed of 4 carbon vacancies ($4V_C$) with a He/V ratio of 1.5, (f) Model of a He-rich platelet precursor embedded between two atomic planes, containing 5 carbon vacancies ($5V_C$) and 12 He atoms (He/V > 2), representing a highly pressurized defect structure. In-plane strain, single atom inserted as substitution and/or interstitial in 6H-SiC bicrystal (g-i) in-plane strain relief ($\varepsilon_{xx}$), (j-l)$\varepsilon_{yy}$.

To further understand the mechanical impact of point defect trapping at GBs, we evaluated local lattice distortions through DFT simulations. In general, we noticed that the segregation of He atom at the GB causes a reduction of atomic local strain compared to the perfect GB. For instance, when a He atom is trapped at an interstitial site within the GB (Fig. 9 i, l), the average strain in the fourth atomic plane from the interface decreases by approximately 0.18% along the x-direction and 21% along the y-direction, relative to a defect-free system. Similar strain-relief behavior was observed when He replaced either a silicon or carbon atom at the GB, as illustrated in Fig. 10 g, h, j and 10k. In addition, the presence of silicon or carbon atoms at the GB reduced the out-of-plane strain by up to 1.5%. These results indicate that the observed strain relief near the GB is not solely due to He depletion (i.e., the denuded zone), but also due to the strong tendency of Si and C atoms to segregate to the boundary, where they help accommodate lattice distortion. This finding is consistent with our experimental observations in Fig. 8, which show reduced strain levels in the vicinity of GBs following He irradiation. This strain reduction highlights the role of GBs not only as physical sinks for defects but also as local strain relaxers, aiding the overall stability of the irradiated microstructure.

In contrast, regions away from the GB exhibit significant strain buildup due to the formation of He-induced bubbles and platelets within the SiC matrix (Fig. 6a), resulting in tensile out-of-plane strain, as shown in Fig. 7. To quantify the contribution of helium-related defects to lattice expansion within grain interiors, we developed a single-crystal 6H-SiC computational model, introducing three distinct representative helium-containing defect configurations (depicted in Fig. 9): (i) a small helium-vacancy cluster composed of four carbon vacancies and two helium atoms (He/V ratio = 0.5) (Fig. 9d), (ii) a larger cluster with increased helium content consisting of four carbon vacancies and six helium atoms (He/V ratio = 1.5) (Fig. 9e), and (iii) a highly pressurized helium platelet precursor containing twelve helium atoms trapped between two atomic planes accompanied by five carbon vacancies (He/V > 2) (Fig. 9f) [71]. After structural relaxation, we calculated the changes in atomic plane spacing relative to a perfect 6H-SiC supercell. All configurations resulted in measurable out-of-plane tensile strain consistent with experimental strain maps obtained via N-PED (Fig. 7). These findings confirm that helium clustering within grain interiors is a key driver of irradiation-induced volumetric swelling and tensile strain. While denuded zones are often used to assess GB sink efficiency, relying solely on their average width provides a limited picture of the complex defect-GB interactions. Nanoscale strain mapping using N-PED offers a more detailed and spatially resolved view, capturing localized strain variations that reflect underlying defect migration, trapping, and relaxation processes. Our N-PED results reveal distinct strain behavior at GBs compared to grain interiors, highlighting the dynamic role of GBs as defect sinks. By attracting and redistributing He atoms and interstitials, GBs reduce defect buildup in the surrounding matrix and alleviate associated lattice strain. This process effectively releases stored elastic energy, helping to mitigate swelling, cracking, and other irradiation-induced degradation mechanisms. These findings underscore the critical influence of GB structure and chemistry on the mechanical response of irradiated ceramics. A more atomistic understanding of strain evolution and defect accommodation at GBs can directly inform the design of radiation-tolerant materials for next-generation nuclear energy systems.

Based on the DFT results, C, Si, and He atoms exhibit a strong energetic preference to segregate at the grain boundary (GB) core in 6H-SiC. Their presence is therefore expected to significantly influence both the chemical environment and the structural integrity of the boundary, with direct implications for the material's mechanical behavior. To quantify these effects, first-principles computational tensile tests (FPCTT) were conducted on various 6H-SiC grain boundary configurations, including pristine (clean), He-, Si-, and C-decorated symmetrical tilt grain boundaries (STGBs), as well as a single-crystal (SC) reference system.

Fig. 10 presents the stress–strain curves for each case. Interestingly, the STGB configuration demonstrates reduced mechanical performance compared to the SC: the SC exhibits a tensile strength of ~16.5 GPa and a fracture strain of 20%, while the clean STGB displays lower values,~10.5 GPa and 14%, respectively, suggesting that the GB acts as a mechanical softening site in this context [72]. This finding suggests that the presence of GB leads to the weakening of the strength (loss of ductility) of SC-SiC.

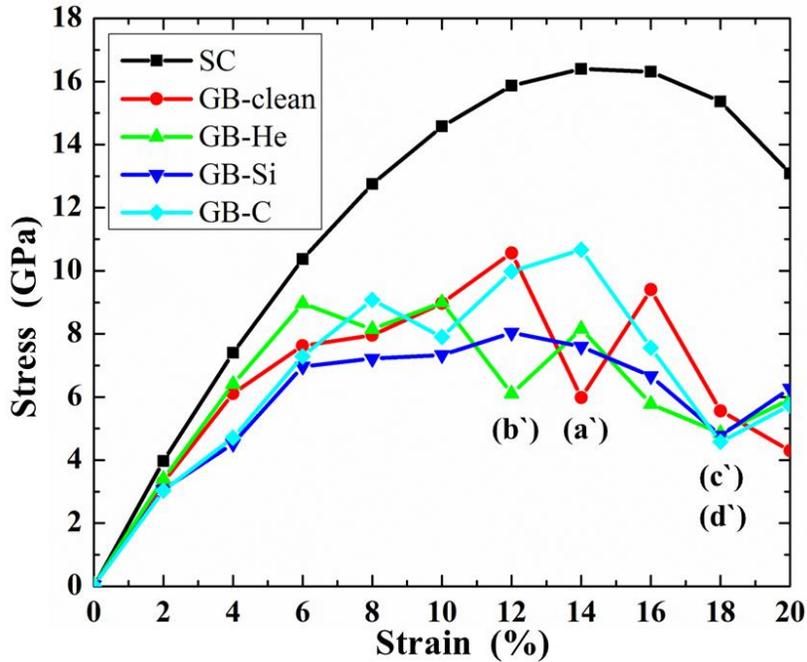

Fig. 10. Stress-strain curves of single-crystal (SC) and grain boundary (GB) configurations of 6H-SiC obtained from first-principles tensile tests. The SC reference shows the highest tensile strength (~16.5 GPa) and fracture strain (~20%). In contrast, the clean GB exhibits reduced strength (~10.5 GPa) and ductility (~14%), reflecting mechanical softening at the interface. When decorated with interstitial He, the GB shows further loss of strength (~9 GPa) and ductility (~12%), indicating strong embrittlement. In comparison, GBs containing Si or C interstitials demonstrate enhanced ductility (by ~28% relative to the clean GB) and comparable or slightly higher strength, highlighting the role of Si and C in reinforcing GB cohesion under tensile loading.

For GBs containing interstitial defects, three distinct behaviors are observed. When He atoms are trapped at the GB, both the tensile strength (~9 GPa) and ductility (fracture strain ~12%) are further reduced relative to the clean GB, indicating a strong embrittling effect. In contrast, the addition of C and Si interstitials leads to a noticeable improvement in ductility, by approximately 28% relative to the clean GB, while maintaining comparable or slightly enhanced strength. These trends suggest that while He segregation degrades GB cohesion, self-interstitials such as Si and C enhance GB stability, likely by reinforcing local bonding networks. Therefore, optimizing GB sink efficiency under irradiation may require concurrent trapping of Si or C interstitials to counteract He-induced embrittlement and prevent premature fracture. Figure 11 illustrates the unstrained (0%) (a-d) and onset of fracture for each configuration (a`-d`). In all cases, fracture initiates along the GB due to bond stretching or breaking, affirming the GB as a preferential failure path. A notable exception is observed in the $I_C$ case, where the GB reaches a maximum tensile stress comparable to that of the clean GB. Detailed structural analysis reveals that the $I_C$ atom tends to attract a neighboring C atom, forming a stable dumbbell configuration (Fig. 11d), which contributes to enhanced local bonding. This behavior is consistent with recent DFT studies suggesting that C interstitials play a key role in healing irradiation-induced damage in 6H-SiC. However, the underlying mechanism of this recovery effect remains unclear and warrants further investigation [73]. To gain deeper insight into the origin of the contrasting mechanical effects of He, Si, and C at GBs, a Bader charge density analysis was performed, as discussed in the following section and shown in Fig. 11.

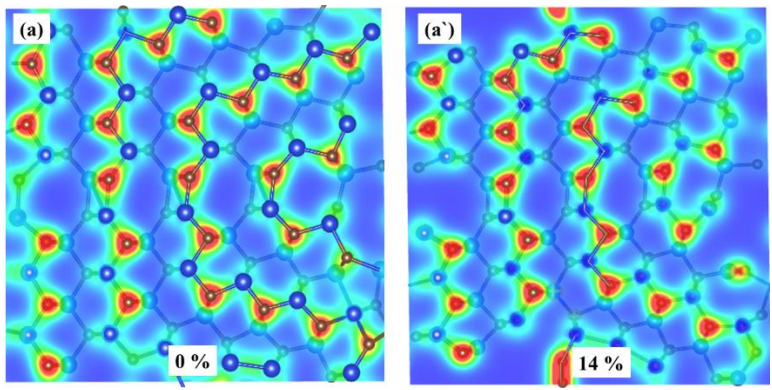

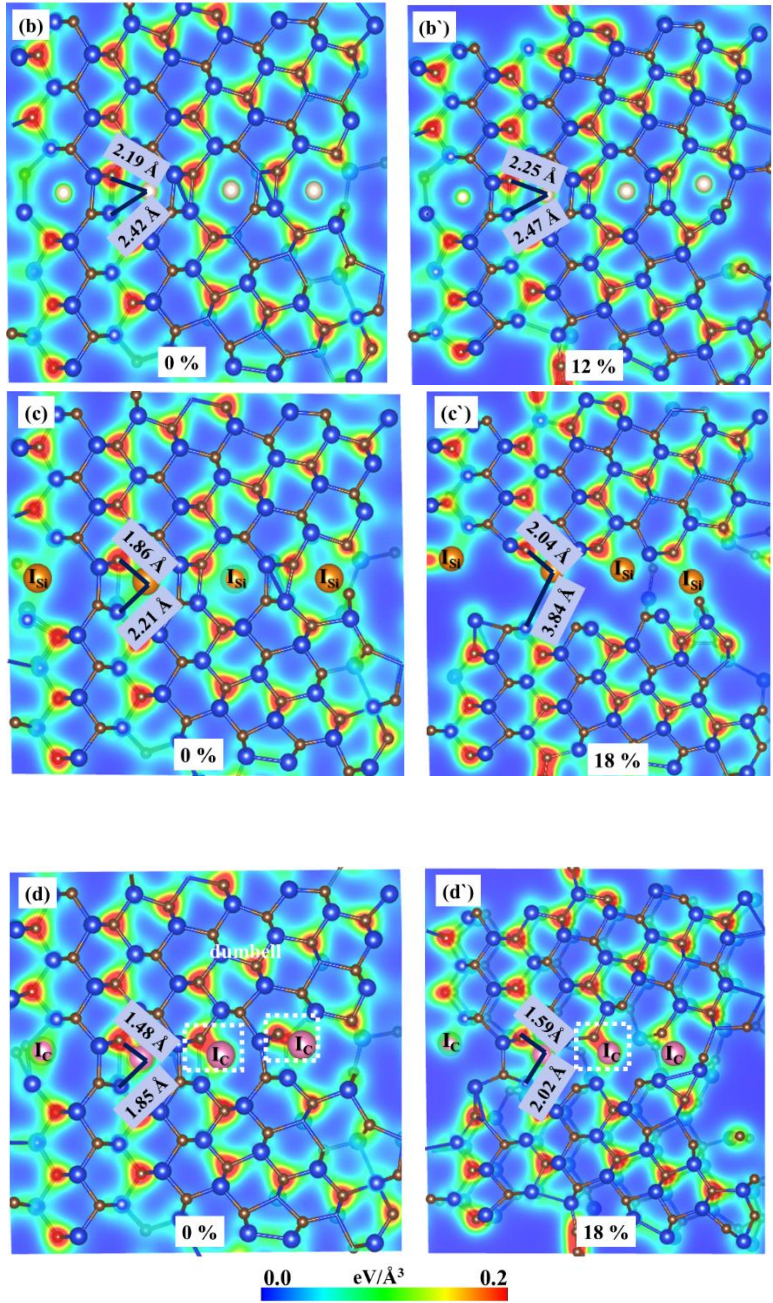

**Fig. 11.** Bader charge density distributions for the clean grain boundary (GB), and GBs containing interstitial He, Si, and C atoms. Panels (a-d) correspond to the unstrained state (0% strain), while panels (a′-d′) show the respective configurations at fracture strain. In the clean GB (a, a′), charge density is relatively uniform with strong localization around C atoms. For the He-decorated GB (b, b′), charge density appears spherical and weakly interacting, reflecting the non-bonding nature of He. In contrast, the Si- and C-decorated GBs (c, c′; d, d′) display anisotropic charge distributions

overlapping with neighboring atoms, indicating enhanced covalent bonding that strengthens local GB cohesion.

To further investigate the origin of the strengthening or weakening effects imparted by different interstitial species at the GB, we performed Bader charge density analysis [74] to assess the redistribution of electronic charge around each atom. Figure 11 presents charge density contour plots for the clean GB (Fig. 11 a) and GBs containing interstitial He, Si, and C atoms (Fig. 11 b–d). In the clean GB, the charge density is relatively uniform around Si and C atoms, with higher electron density localized near C atoms, consistent with previous studies showing that bonding in SiC is predominantly covalent ($sp^3$ hybridized) with a minor ionic contribution arising from the electronegativity difference between Si and C atoms [75]. The introduction of interstitial species at the GB core modifies the local charge distribution. As shown in Fig. 11b, the He interstitial exhibits a spherical and symmetric charge density, with minimal overlap with neighboring atoms. This configuration suggests a non-bonding, weakly interacting (ionic-like) nature, resulting in weakened local bonding and mechanical embrittlement. In contrast, interstitial Si and C atoms (Figs. 11c and 11d) exhibit anisotropic charge distributions that overlap with neighboring atoms, indicative of enhanced covalent bonding at the GB. The nature of bonding between different interstitials and one C atom of their nearest neighbor atoms can be further investigated by analyzing their projected density of states (PDOS) on both S and P orbitals.

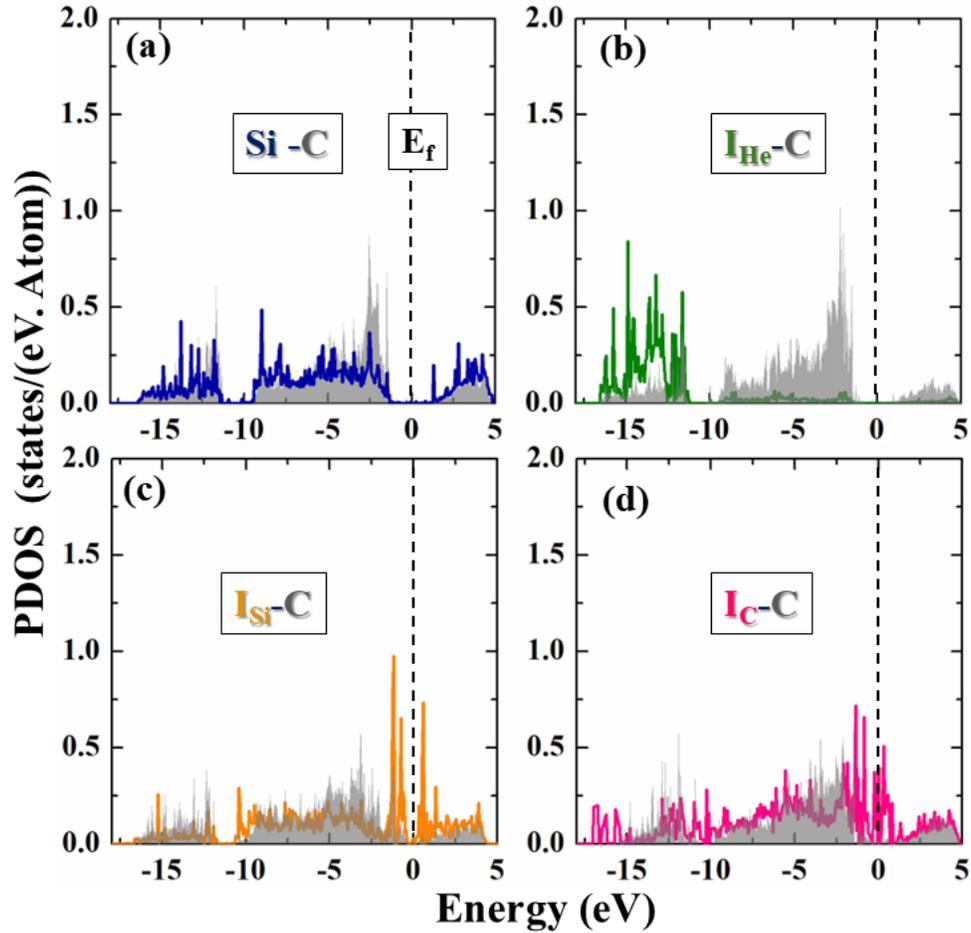

**Fig. 12.** Atom-projected density of states (PDOS) for different interstitial atoms with one of the 1st NN C atom (a) Si in clean GB, (b) I$_{He}$ and 1st NN C atom, (c) Isi and 1st NN C atom, and (d) I$_C$ and 1st NN C atom in GB 6H-SiC. A vertical dashed line marks the position of the Fermi level at 0 eV energy.

To complement the charge analysis, we calculated the PDOS for the clean and defect-containing GBs (Fig. 12). In the clean GB (Fig. 12a), the PDOS of nearest-neighbor (NN) Si and C atoms shows strong orbital overlap, with a distinct ~2 eV band gap between the occupied and unoccupied states, consistent with semiconducting behavior and robust sp$^3$ bonding, as also reported in prior DFT studies [75–77]. In the case of He interstitials (Fig. 12b), the PDOS of He and its 1st NN C atom shows no hybridization, and the He 1s electrons reside entirely in the occupied states, forming a sharp, isolated peak. This aligns with the charge density analysis, further confirming that He disrupts local bonding by failing to form meaningful electronic interactions with neighboring atoms. Conversely, both Si and C interstitials (Figs. 12c and 12d) show substantial orbital overlap with adjacent C atoms in the PDOS, particularly near the Fermi level. This overlap suggests the formation of metallic-like bonds, especially pronounced for the $I_C$, which induces a significant increase in electronic states at the Fermi level. Structural analysis reveals that the $I_C$ atom forms a dumbbell configuration by bonding with a 1$^{st}$ NN C atom, resulting in a localized accumulation of charge. This geometry is consistent with sp hybridization and the formation of a double bond between the two C atoms. Together, the Bader charge and PDOS analyses reveal that while He atoms weaken the GB by disrupting local bonding, C and Si interstitials enhance GB cohesion through the formation of covalent and, in some cases, metallic bonds. The improved mechanical performance observed with C interstitials is particularly linked to the presence of sp- and sp$^3$-type bonding networks, which contribute to the increased strength and ductility of the GB.

## 4. Conclusions

This study demonstrates the power of combining high-resolution nanoscale strain mapping via nano-beam precession electron diffraction (N-PED) with atomistic modeling to understand irradiation-induced deformation in silicon carbide. In He and H ion-irradiated single-crystal 4H-SiC, we first validated N-PED strain profiles against high-resolution X-ray diffraction (HR-XRD) simulations, confirming peak out-of-plane tensile strain of ~1.15% at the projected ion range and revealing anisotropic strain distributions. Building on this validated framework, we applied N-PED to study polycrystalline α-SiC, where conventional techniques fail to capture local strain effects. The results uncovered significant tensile strain at grain boundaries (up to ~2.5%) and simultaneous strain relief in adjacent regions, indicating GBs act as efficient defect sinks. These experimental observations were corroborated by DFT calculations, which revealed reduced migration barriers for C and Si interstitials and helium atoms near GBs, promoting defect segregation and localized strain relaxation. Simulations of He-containing defect structures further confirmed that He clusters contribute to tensile strain in grain interiors, while GBs mediate strain relief through defect absorption. First-principles computational tensile tests confirm that Si and C interstitials enhance GB cohesion, whereas He induces GB embrittlement. Charge density distributions reveal weak interactions between He atoms and their neighboring atoms, in contrast to the anisotropic and directional bonding of Si and C. Density of states (DOS) analyses further show that the hybridization between s and p orbitals is a key factor in determining bond type and strength. Collectively, the experimental and atomistic results provide detailed insight into the effects of irradiation-induced point defects on the mechanical behavior of SiC and highlight the critical role of grain boundary engineering in improving the radiation tolerance of ceramic materials for advanced nuclear and space applications.

**Acknowledgements**

This work was financially supported by the European Union under the project Robotics and advanced industrial production (Reg. No. CZ.02.01.01/00/22_008/0004590). CzechNanoLab project LM2023051 funded by MEYS CR is gratefully acknowledged for the financial support of the measurements/sample fabrication at LNSM Research Infrastructure. Parts of this research were carried out at IBC at the Helmholtz-Zentrum Dresden-Rossendorf e. V., a member of the Helmholtz Association. The author A. M. acknowledge the assistance provided by the Ferroic Multifunctionalities project, supported by the Ministry of Education, Youth, and Sports of the Czech Republic. Project No. CZ.02.01.01/00/22_008/0004591, co-funded by the European Union. CANAM infrastructure was used to provide RBS/C measurements. This work was supported by the Ministry of Education, Youth and Sports of the Czech Republic through the e-INFRA CZ (ID:90254)

# Nanoscale Strain Evolution and Grain Boundary-Mediated Defect Sink Behavior in Irradiated SiC: Insights from N-PED and DFT

N. Daghbouj[a*], A.T. AlMotasem[a*], J.Duchoň[b], B.S. Li[c], M. Bensalem[a], F. Munnik [d], Xin Ou[e], A. Macková[f], W.J. Weber[g], T. Polcar[a,h]

[a]*Department of Control Engineering, Faculty of Electrical Engineering, Czech Technical University in Prague, Technická 2, 160 00 Prague 6, Czechia*

[b]*Institute of Physics of the Czech Academy of Sciences, Na Slovance 1999/2, 182 21 Prague 8, Czechia*

[c]*State Key Laboratory for Environment-friendly Energy Materials, Southwest University and Technology, Mianyang, Sichuan 621010, China*

[d]*Helmholtz–Zentrum Dresden–Rossendorf, Institute of Ion Beam Physics and Materials Research, Bautzner Landstr. 400, 01328 Dresden, Germany*

[e]*State Key Laboratory of Functional Materials for Informatics, Shanghai Institute of Microsystem and Information Technology, Chinese Academy of Sciences, 865 Changning Road, Shanghai 200050, China*

[f]*Nuclear Physics Institute of the Czech Academy of Sciences, 250 68 Husinec-Řež, Czechia*

[g]*Department of Materials Science & Engineering, University of Tennessee, Knoxville, TN 37996, USA*

[h]*School of Engineering, University of Southampton, Southampton SO17 1BJ, United Kingdom*

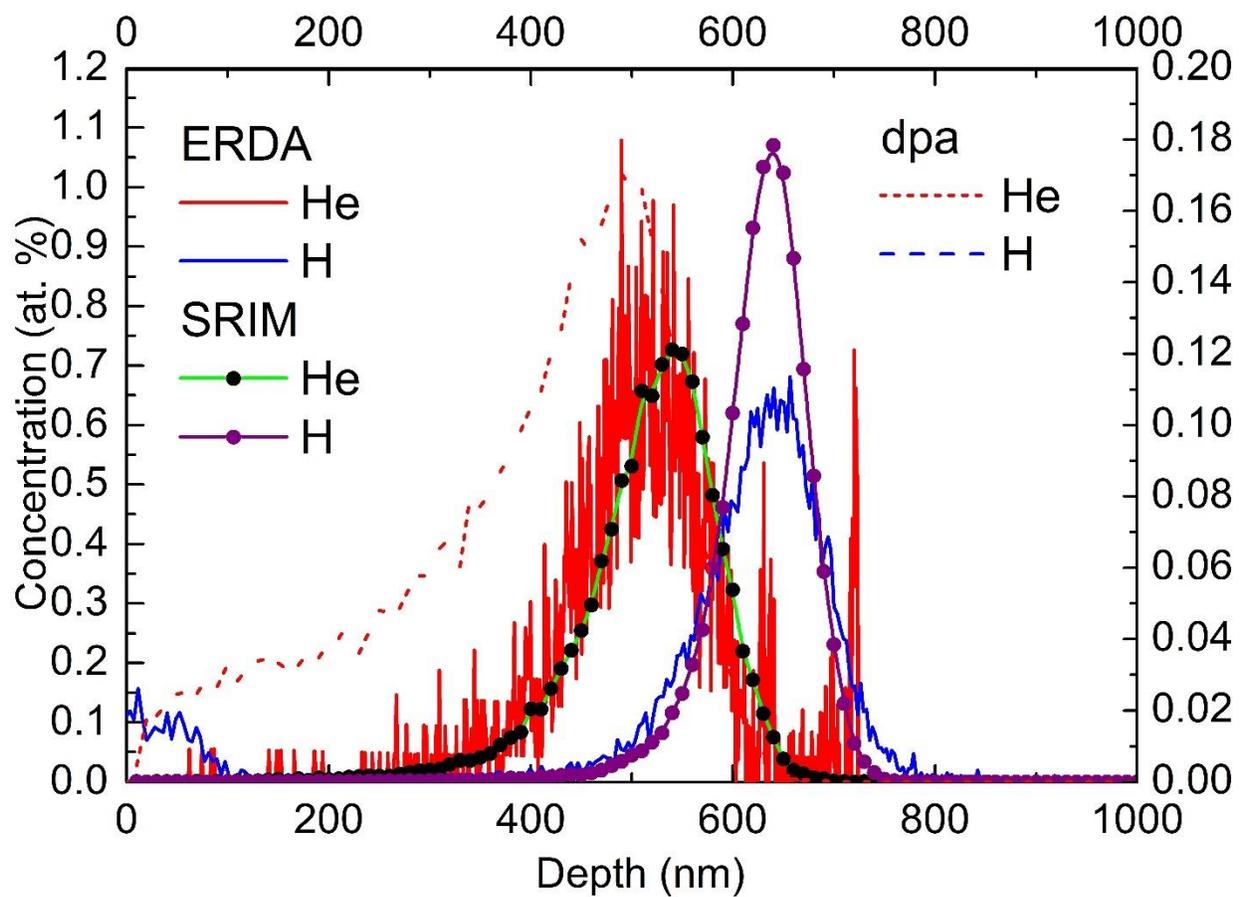

Figure S1. He and H concentration depth profiles were calculated using the SRIM calculation and ERDA experiment in the sample implanted by $1 \times 10^{16}$ ions/cm$^2$.

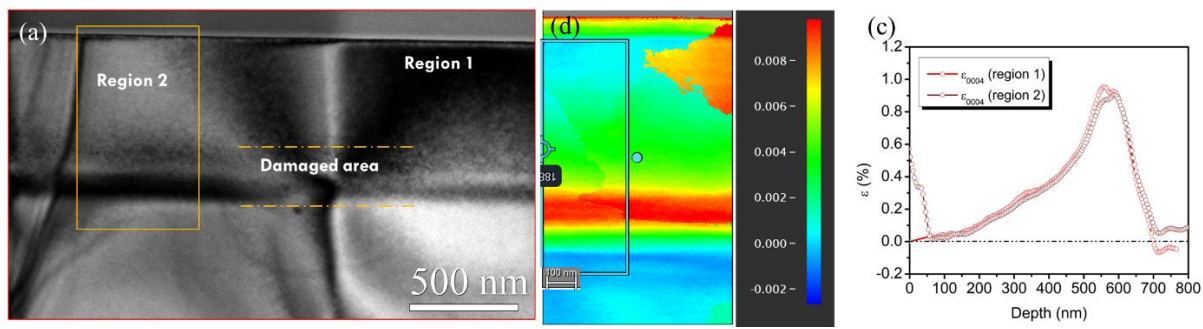

Figure S2. (a) Bright-field transmission electron microscopy (BF-TEM) image of 4H-SiC co-implanted with He and H ions (fluence: $1 \times 10^{16}$ ions/cm²), oriented along the [110] zone axis. (b) Strain map of $\varepsilon_{0004}$ acquired using nano-beam precession electron diffraction (N-PED) in the region 2 highlighted by the yellow square in (a). (c) Comparison of the depth-dependent strain profile obtained from TEM-based N-PED mapping region 1 (wine) and TEM-based N-PED mapping region 1(red).

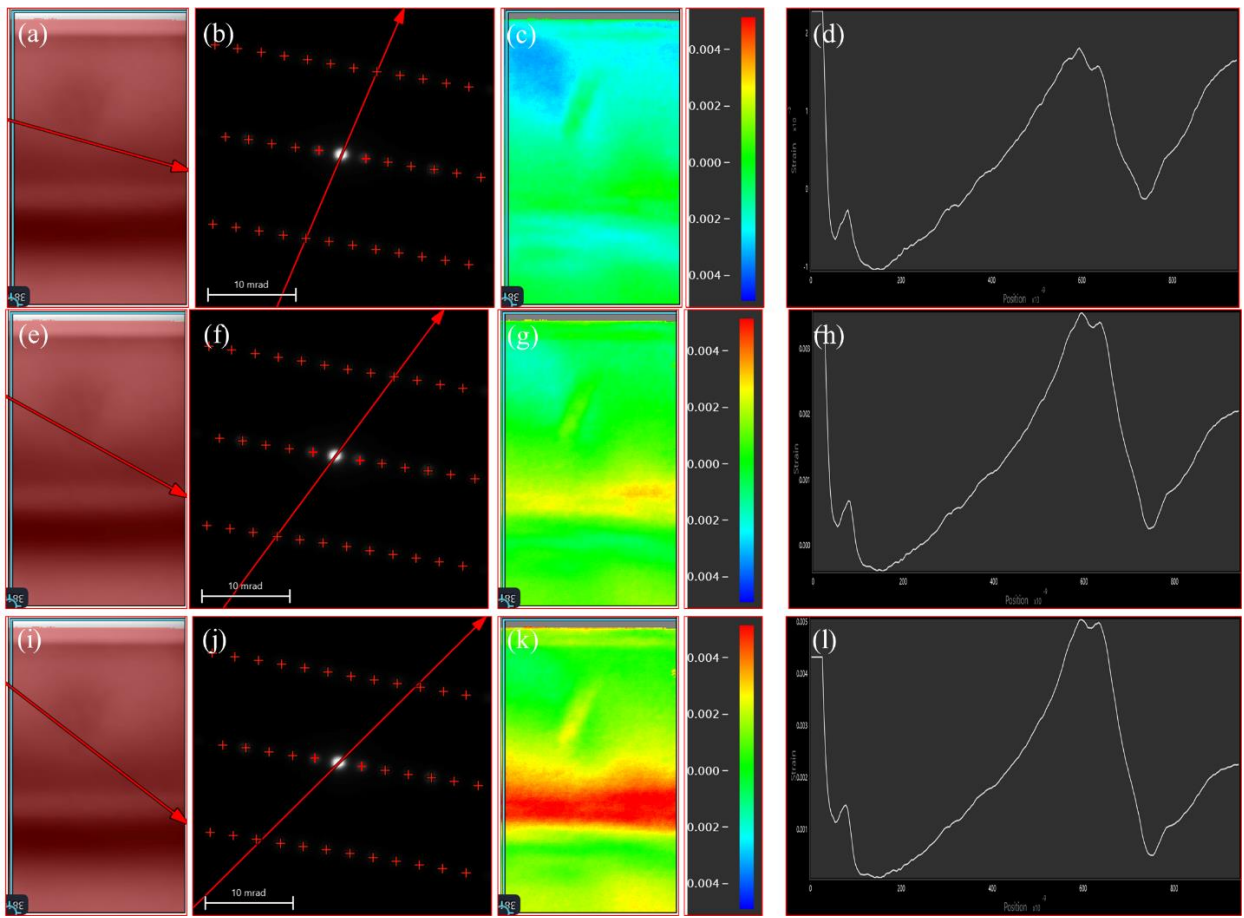

Figure S3. (a, e, i) Virtual bright-field transmission electron microscopy (VBF-TEM) images of the He-H co-implanted 4H-SiC sample, oriented along the [110] zone axis.
Strain Analysis along oblique crystallographic directions [1-10n]: (b–d) Strain along [1-102]: (b) Selected-area diffraction pattern (SADP) indicating the measurement direction; (c) N-PED strain map showing the $\varepsilon_{1\text{-}102}$ component, primarily revealing compressive strain; (d) corresponding strain profile. (f–h) Strain along [1-104]: (f) SADP indicating the measurement direction; (g) N-PED strain map for $\varepsilon_{1\text{-}104}$, showing tensile strain; (h) associated strain profile. (j-l) Strain along [1-106]: (j) SADP indicating the measurement direction; (k) N-PED strain map for $\varepsilon_{1\text{-}106}$, revealing tensile deformation; (l) corresponding strain profile.

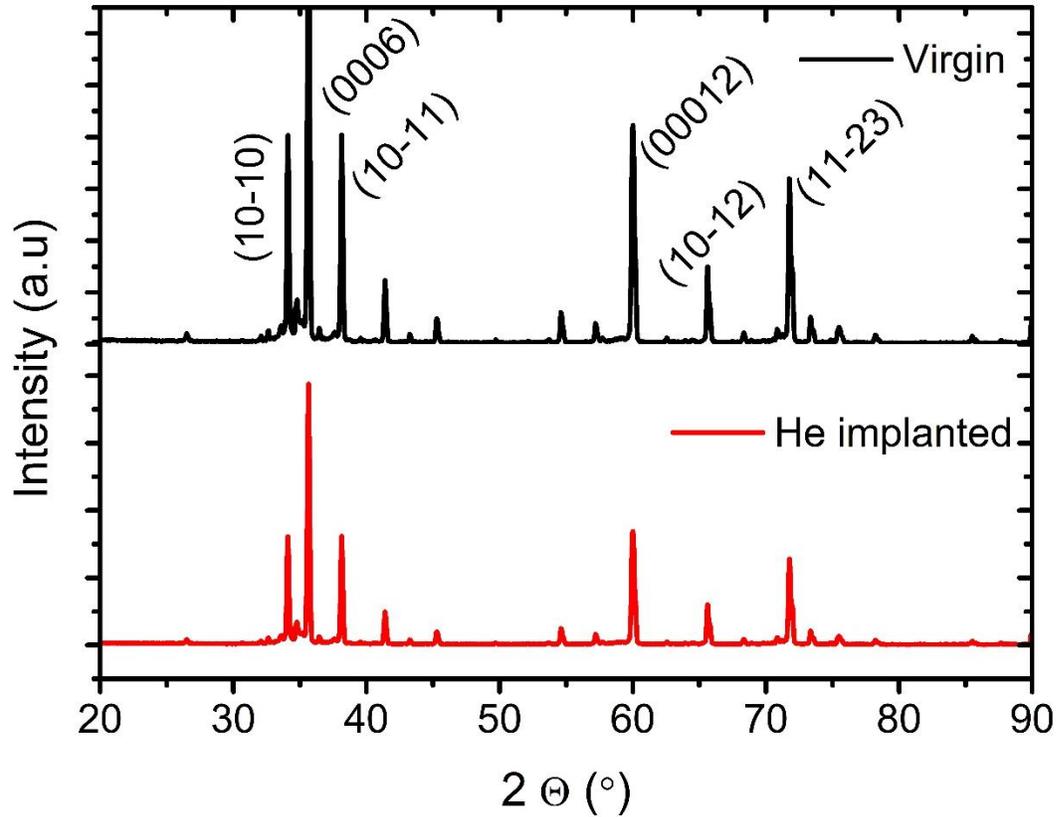

Figure S4. X-ray diffraction of virgin and He irradiated sintered SiC.

Figure S5 presents detailed TEM and HRTEM analyses of helium-irradiated polycrystalline SiC to directly visualize irradiation-induced defects. Bright-field TEM under over- and under-focus conditions reveals helium-related defects as contrasting features, while high-magnification imaging confirms the presence of nanometer-scale bubbles and extended platelets at the depth corresponding to the peak helium concentration (~900 nm). HRTEM further resolves platelet structures and demonstrates that, despite significant helium accumulation, grain boundaries remain crystalline rather than amorphous. This provides direct microstructural evidence for helium defect formation and the role of grain boundaries as defect sinks.

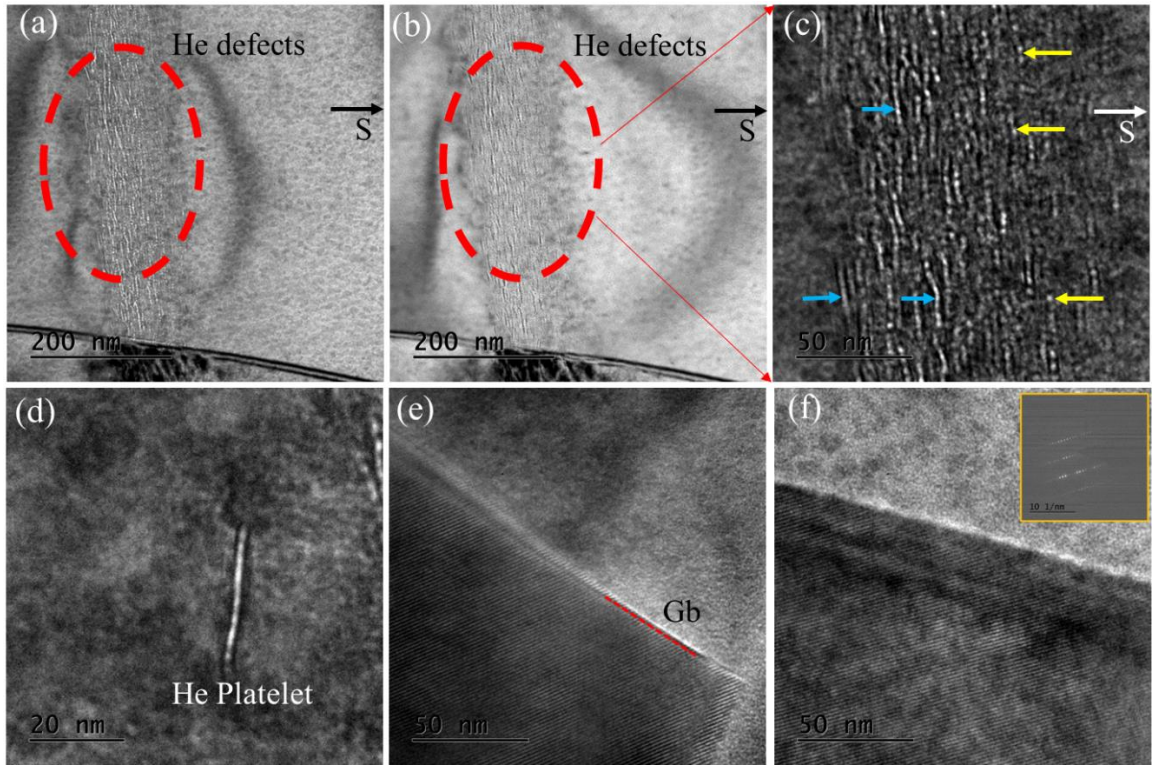

Fig.S5. Helium irradiation effects in polycrystalline SiC (800°C, $1\times10^{17}$ He/cm²). (a) Over-focused bright-field TEM image showing helium-induced defects as white contrasts. (b) Under-focused bright-field TEM image of the same region, where defects appear in black. (c) High-magnification TEM of the peak He concentration region (indicated by dashed red circles in panels a and b, imaged under over-focus conditions), highlighting helium bubbles (yellow arrows) and helium platelets (cyan arrows). (d) HRTEM image of a helium platelet. (e, f) HRTEM images of a grain boundary at and near the peak He concentration, confirming the preservation of crystallinity. "S" indicates the surface direction.

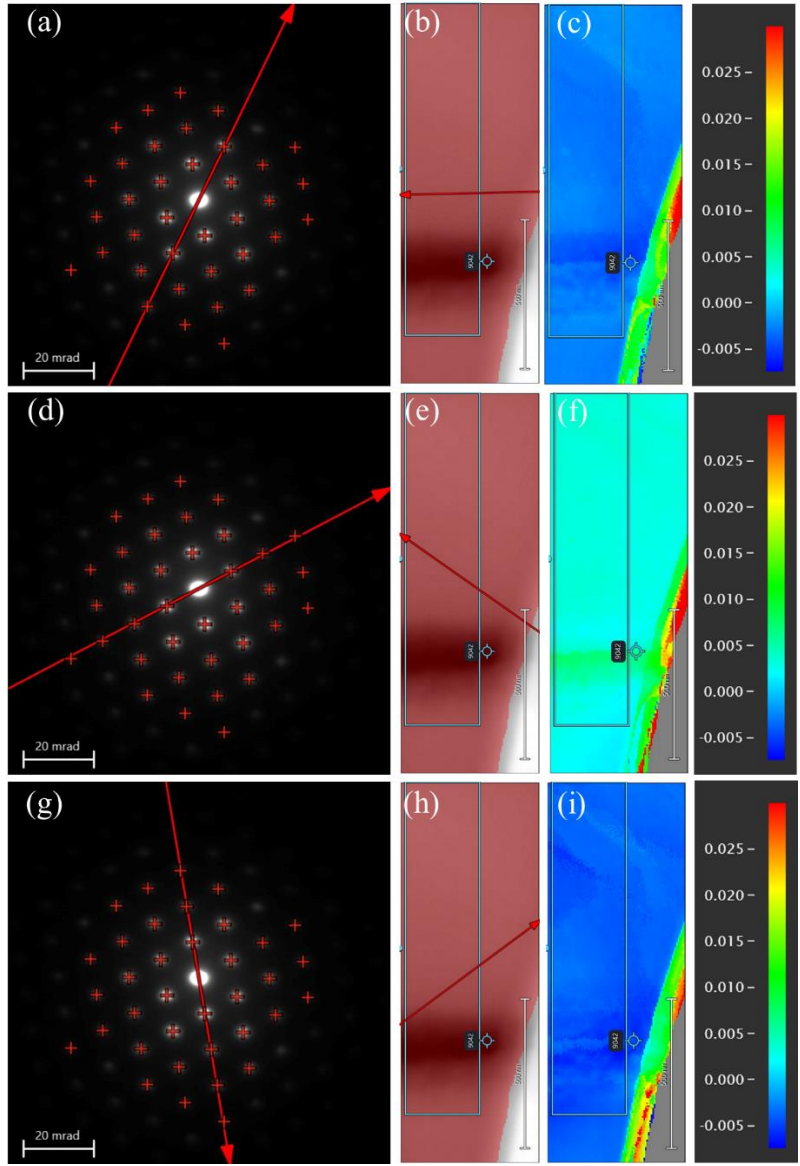

Figure S5. Strain Analysis along various crystallographic directions in the He-implanted sintered SiC sample with $1\times10^{17}$ He/cm$^2$, oriented along the [241] zone axis. (a-c) Strain along [2-1-10], (a) Selected-area diffraction pattern (SADP) indicating the measurement direction; (b) Virtual bright-field transmission electron microscopy (VBF-TEM) images indicating the measurement direction of the strain, (c) N-PED strain map showing the $\varepsilon_{2\text{-}1\text{-}10}$ component, revealing compressive strain; (d-f) Strain along [10-1-2], (d) Selected-area diffraction pattern (SADP) indicating the measurement direction; (e) Virtual bright-field transmission electron microscopy (VBF-TEM) images indicating the measurement direction of the strain, (f) N-PED strain map showing the $\varepsilon_{10\text{-}1\text{-}2}$ component, revealing small tensile strain and peaked at peak He concentration

in the region far from the GB; (g-i) Strain along [1-102], (g) Selected-area diffraction pattern (SADP) indicating the measurement direction; (h) Virtual bright-field transmission electron microscopy (VBF-TEM) images indicating the measurement direction of the strain, (i) N-PED strain map showing the $\varepsilon_{1\text{-}102}$ component, revealing compressive strain and peaked at peak He concentration in the region far from the GB;